%% file: NIST-analytics.tex
\PassOptionsToPackage{table}{xcolor}

\documentclass[sigconf,preprint]{acmart}
\pdfoutput=1
\hypersetup{
    pdftitle = {\@title},
    pdfauthor = {\@author}
}

\usepackage{tikz}

\usepackage{caption}
\usepackage[T1]{fontenc}
\usepackage[latin1]{inputenc}
\usepackage{fancyvrb}
\usepackage{subcaption}
\usepackage{graphicx}

\graphicspath{.}

\usepackage{todonotes}
\newcommand{\TODO}[1]{\todo[inline]{#1}}

\newcommand{\FILE}[1]{}

\newcommand{\OK}{+}
\newcommand{\OP}{o}
\newcommand{\NA}{-}

\definecolor{lightgray}{gray}{0.94}
\let\oldtabular\tabular
\let\endoldtabular\endtabular
\renewenvironment{tabular}{\rowcolors{2}{lightgray}{white}\oldtabular}{\endoldtabular}

\setcopyright{none}
\copyrightyear{2022, 2023}
\acmYear{2022, updated Oct 2023}
\acmDOI{}

\acmConference[White Paper]{White paper produced by the NIST Public Working Group on Big Data}{White paper produced by the NIST Public Working Group on Big Data}{Washington, D.C.}
\acmBooktitle{NIST Public Working Group on Big Data, Washington, D.C.}
\acmPrice{}
\acmISBN{}

\makeatletter
\def\@copyrightspace{\relax}
\makeatother

\begin{document}

\newpage

\title{Whitepaper on Reusable Hybrid and Multi-Cloud Analytics Service Framework}

\input{section/abstract}

\maketitle

\tableofcontents
\newpage

\input{authors}



\newpage

\maketitle

\input{section/summary}

\input{section/concepts}

\input{section/technologies}

\input{section/usecases}



\input{usecase/hvac}

\input{usecase/security}

\input{usecase/earthquake}

\input{section/architecture}

\input{section/security}

\section{Implementation}

\input{section/defining}

\input{section/catalog}

\input{section/registry}

\input{section/federation}

\input{section/workflow}

\input{section/data}

\input{section/package}

\section{Application}

Next, we list some elementary application examples that identify the
usefulness of this framework.

\input{section/experimment}

\input{section/nlp}

\input{section/eq-result}

\input{section/conclusion}

\input{NIST-acknowledgement}

\bibliographystyle{ACM-Reference-Format}
\bibliography{NIST-analytics}



\end{document}

%% file: section/abstract.tex
\begin{abstract}

\FILE{section-abstract.tex}

Over the last several years, the computation landscape for conducting
data analytics has completely changed. While in the past, a lot of the
activities have been undertaken in isolation by companies, and
research institutions, today's infrastructure constitutes a wealth of
services offered by a variety of providers that offer opportunities
for reuse, and interactions while leveraging service collaboration,
and service cooperation.

This document focuses on expanding analytics services to develop a
framework for reusable hybrid multi-service data analytics. It
includes (a) a short technology review that explicitly targets the
intersection of hybrid multi-provider analytics services, (b) a small
motivation based on use cases we looked at, (c) enhancing the concepts
of services to showcase how hybrid, as well as multi-provider services
can be integrated and reused via the proposed framework, (d) address
analytics service composition, and (e) integrate container
technologies to achieve state-of-the-art analytics service deployment
capabilities.

\end{abstract}

%% file: authors.tex
\author{Gregor von Laszewski}
\email{laszewski@gmail.com}
\orcid{0000-0001-9558-179X}
\affiliation{%
  \institution{University of Virginia}
  \streetaddress{Biocomplexity Institute\\
                Town Center Four\\
                994 Research Park Boulevard}
  \city{Charlottesville}
  \state{VA}
  \postcode{22911}
  \country{USA}
}

\author{Wo Chang}
\email{wchang@nist.gov}
\affiliation{%
  \institution{NIST}
  \streetaddress{100 Bureau Drive}
  \city{Gaithersburg}
  \state{MD}
  \postcode{20899}
  \country{USA}
}

\author{Russell Reinsch}
\email{russell@mcleansystemsintegration.com}
\affiliation{%
  \institution{McLean Systems Integration}
  \streetaddress{6102 Franklin Park}
  \city{McLean}
  \state{VA}
  \postcode{22101}
  \country{USA}
}

\author{Olivera Kotevska}
\email{kotevskao@ornl.gov}
\affiliation{%
  \institution{Oak Ridge National Laboratory}
  \city{Oak Ridge}
  \state{TN}
  \postcode{37830}
  \country{USA}
}

\author{Ali Karimi}
\email{ali.karimi@ieee.org}
\affiliation{%
  \institution{Trans Technology Global}
  \streetaddress{2001 Wilshire Blvd Suite 515}
  \city{Los Angeles}
  \state{CA}
  \postcode{90403}
  \country{USA}
}

\author{Abdul Rahman Sattar}
\email{abdul.sattar@arcticwolf.com}
\orcid{0000-0003-1598-4674}
\affiliation{%
  \institution{Arctic Wolf Networks}
  \city{Eden Prairie}
  \state{MN}
  \country{USA}
}

\author{Garry Mazzaferro}
\email{garymazzaferro@gmail.com}
\affiliation{%
  \country{USA}
}

\author{Geoffrey C. Fox}
\email{gcfexchange@gmail.com}
\affiliation{%
  \institution{University of Virginia}
  \streetaddress{Biocomplexity Institute\\
                Town Center Four\\
                994 Research Park Boulevard}
  \city{Charlottesville}
  \state{VA}
  \postcode{22911}
  \country{USA}
}

\renewcommand{\shortauthors}{G. von Laszewski, et al.}

%% file: section/summary.tex
\FILE{section-summary}

section/data.tex
	section/experimment.tex
	section/registry.tex
	usecase/earthquake.tex
\section{Introduction}
\label{sec:summary}

Analytics as a service has become a multi-billion dollar opportunity for the industry. With Big Data's compound annual growth rate at 61\% and its ever-increasing deluge of information in the
mainstream, the collective sum of world data will grow from 33
zettabytes (ZB, 1021) in 2018 to 175 ZB by 2025 \cite{www-idc-forecast}.
The presence of such a rich source
of information requires a massive analysis that can effectively bring
about much insight and knowledge discovery. While in the past, emphasis has been placed on the hybrid multi-cloud {\em infrastructure
services}, the focus is now significantly shifting to offering {\em
analytics as services} and not just infrastructure as a service to
customers and researchers. Hence, it is important to identify how
researchers and industry can interoperate with services offered by
various service providers. Analog to the terminology in cloud
computing, we introduce the terms {\em hybrid} analytics services to
include services run by remote service providers or by an organization
on local resources. In addition, we use the term {\em multi}
analytics services to indicate that multiple services from potentially
multiple service providers work in concert to offer a new capability
released to users as analytics services. While applying such services
to data, we term the combination of such services as {\em hybrid
multi-services data analytics} services. If properly put in place, the
resulting service accelerates new solutions offered by industry and
research as new services with a combined new functionality, which can
be {\em reused} or {\em replicated}. Research institutions and private
companies offer a number of services that can be integrated into
custom analytics services, offering {\em competing} but also {\em
collaborating} analytics services.

To achieve this goal of developing and integrating such services, a certain degree of platform-independence and
platform interoperability is needed to ensure that the pathway to
leverage these hybrid and multi-analytics services are kept at a high
level while at the same time exposing enough details. 
It is advantageous to leverage frameworks that are used by many vendors. Here
we will use REST services as such a platform as it provides us with
the needed abstractions but also allows us to integrate with persistent services such as data services.

Furthermore, we need to support intelligent decision-making as part of
the service orchestration. Services must be chosen to fulfill a set of
{\em analytics service} level requirements posed by the users. It is
of particular interest how we can formulate hybrid analytics services
and multi-analytics services offered by different providers that
provide other features. The user needs to specify this via a simple
analytics service provider independent specification.

Our data analysis intends to be capable of determining which service is
suitable or chosen based on its requirements and to what degree
reusability is offered while replicating the analysis across different
services. Hence we will work towards a {\em ``Reusable Hybrid
Multi-Services Data Analytics Framework''}. This results in a research
platform that allows the creation of an integrated application
platform benefiting from reusable hybrid analytics services.

By integrating such services, we will be able to significantly impact
data analytics while leveraging not only one vendor's implementation
but by promoting the reuse services via a {\em many-vendors}
approach. Not only that, but we will also allow the interplay between
different approaches while offering a uniform specification platform.
Because we target the topic of this interplay, the effort has been
done in collaboration with the NIST Information Access Division (IAD)
in the NIST Big Data Working Group.

The paper is structured as follows. First, we present the motivation
leading up to this work (Section~\ref{s:background}), followed by a
discussion about requirements that we derived by analyzing a number of
complex use cases. Next, we present our architectural approach, which is
based on lessons learned from the requirements we have gathered and
lessons learned during our implementation (Section~\ref{s:arch}. We
present an architectural design capable of supporting the needs we
have identified. Finally, we present our conclusions.


%% file: section/concepts.tex
\FILE{section-concepts.tex}

\section{Enabling Concepts}\label{s:background}

This section explains the motivation while briefly summarizing the
different concepts constituting our work. While our previous work
focused on developing a Big Data Reference Architecture and standards
roadmap \cite{nist-v8}. This work specifically focused on the
definition of {\bf\em Analytics Services}. This work is a logical
enhancement to the earlier work and can leverage activities conducted
as part of the NIST Big Data Reference Architecture (NBD-RA) and
NBD-RA Interfaces. However, the work here targets explicitly {\bf\em
Data Analytics} as a pathway to integrate the data analytics
ecosystem. This includes not only existing legacy analytics services
and tools but also the integration of state-of-the-art AI services,
including machine learning and deep learning analytics, within the
auspice to create a service-oriented framework integrating all of
them. Hence, they can easily be reused by others. Next, we define some
of the terminology and concepts we use.

\begin{description}

\item[From Big Data Reference Architecture to Analytics Services.]
\label{s:arch} NIST has developed a Big Data Reference Architecture as part of
NBDIF\cite{nist-v6} and identified a number of use cases that motivate
it \cite{nist-v3}. We leverage this effort
~\cite{nist-v1,nist-v2,nist-v3,nist-v4,nist-v5,nist-v6,nist-v7,nist-v8,nist-v9}
while formulating service interoperability specifications that we
focus on in this effort and have not been previously addressed in
detail. While we previously focused mostly on infrastructure
management, this effort enhances the activities to include high-level
coordinated service deployments and utilization while leveraging
containers. The concepts we introduce next specifically target
analytics services and not just infrastructure services.  However, the
lessons learned from the earlier work significantly influenced this
activity.

\item[Hybrid Analytics Services.]

A {\em hybrid analytics service} combines the strength of analytics
services that are offered by providers in public, private, or
on-premise usage scenarios. It leverages them to provide optimized
orchestration across private, public, and on-premise
analytics. Optimization benefits are not limited to reducing cost but
also addressing security and privacy concerns when the data analytics
or the data to perform the analytics can not be hosted in public
clouds. Many of the major cloud providers such as AWS, Azure, Google,
IBM, Oracle, and others have made hybrid clouds a cornerstone of their
business model, with each of them essentially promoting their own
solutions. Recently, however, we see that the cloud provider's focus
is no longer offering just infrastructure but instead to provide
services hiding and abstracting the cloud infrastructure entirely from
the users while placing focus on offering services. This has led to
vendors also providing hybrid analytics solutions that may integrate
multiple services offered by various providers, resulting in solutions
with heterogeneous service offerings. Integrating such services
involves significant challenges, as each vendor may conform to and or
require different integration solutions for addressing various public,
private and on-premise analytics services. Customers will generally
benefit from a more integrated approach to ease deployment and
management concerns.

\item[Multi-Analytics Services to Cooperate and Compete.]

Over the last several years, we have seen an explosion of analytics
services, mainly through the integration of AI and deep learning
services. High-level analytics services are being developed that hide
and abstract the complex infrastructure needed to embed not only
services from one vendor but multiple vendors. Hence we speak of {\em
multi-analytics services}. These services can then be used in {\em
cooperation} and/or {\em competition}. We cooperate if services
enhance each other, we compete if a service is chosen over another
service due to better service level agreements. Through this interplay
of the services, it is beneficial to formalize interoperability
between them. In cases of competition, we also need to be able to
formulate a competing service that then calls out other services to
implement desired analytics tasks.

\item[Identification of State-of-the-Art Data Analytics Patterns.]

Analytics services consumers have to ask why and how now,
this opportunity can be addressed to enable this interplay by utilizing
hybrid multi-service-based analytics as a service needs are clearly
motivated by state-of-the-art data analytics capabilities that have
only recently became available. In addition, government agencies have
provided some of the most capable high-end computing systems over the
last years, they tightly integrated specialized GPUs as well as
container technologies to bring forward new data analytics
capabilities in these on-premise services. Industry has provided
advanced analysis capabilities for some time but have only recently
reached a maturity supporting reuse and cooperation opportunities
between them.

\item[FAIR Principle for Analytics Services.]
\label{sec:fair}

Reusability is an essential part of adaptation. To make it explicitly
clear, we adopt the well-known FAIR principal \cite{fair-principle} but enhance
them first by focussing on analytics services, deployability, and
operations. Together we use the tern Analytics Service FAIR Principle
(AS-FAIR-DO).

To project easy reusability, we strive toward the implementation of
the AS-FAIR-DO principle for analytics services. The FAIR principle is
typically applied to data; as such, we can apply it to the metadata
associated with analytics services. The FAIR principle addresses which
to be findable, accessible, interoperable, and reusable. In
Figure \ref{fig:as-fair-do} we explicitly augmented the general FAIR
principle with terminology so it can apply to analytics services. The
augmentations are colored in red. As such, not only data is part of the
principle, but also the data representing the services themselves.

\end{description}

\begin{figure*}[htb]
\centering\resizebox{1.0\columnwidth}{!}{
\begin{tabular}{p{1cm}p{12cm}}
\multicolumn{2}{l}{To be Findable:} \\
F1 & \textcolor{red}{analytics services metadata} are assigned a globally unique and persistent identifier \\ 
F2 & \textcolor{red}{analytics services} data are described with rich metadata (defined by R1) \\
F3 &  \textcolor{red}{analytics services metadata} clearly and explicitly include the identifier of the data related to the analytics services it describes \\ 
F4 & \textcolor{red}{analytics services metadata} are registered or indexed in a searchable resource \\
\multicolumn{2}{l}{To be Accessible:} \\
A.1 &  \textcolor{red}{analytics services metadata} are retrievable by their identifier using a standardized communications protocol \\
    A1.1 & \textcolor{red}{analytics services} the protocol is open, free, and universally implementable \\
    A1.2 & the \textcolor{red}{analytics services} protocol allows for an authentication and authorization procedure, where necessary \\ 
A.2 & \textcolor{red}{metadata} are accessible, even when the data are no longer available \\
\multicolumn{2}{l}{To be Interoperable:}\\
I1. & \textcolor{red}{analytics services metadata} use a formal, accessible, shared, and broadly applicable language for knowledge representation. \\
I2. &  \textcolor{red}{analytics services metadata} use vocabularies that follow FAIR principles \\
I3. &  \textcolor{red}{analytics services metadata} include qualified references to other metadata \\
\multicolumn{2}{l}{To be Reusable:} \\
R1. & \textcolor{red}{analytics services metadata} are richly described with a plurality of accurate and relevant attributes \\
R1.1 & (meta)data are released with a clear and accessible data usage license \\
R1.2 & (meta)data are associated with detailed provenance \\
R1.3 & (meta)data meet domain-relevant community standards \\
\multicolumn{2}{l}{To be Deployable:}\\
D.1 & \textcolor{red}{analytics services metadata describing deployability aspects} \\
\multicolumn{2}{l}{To be Operational:} \\ 
O.1 & \textcolor{red}{analytics services metadata describing operational aspects} \\ 
\end{tabular}
}
\caption{Fair guiding principles adapted to analytics services:
  Analytics Services - FAIR - Deployable and Operational
  (AS-FAIR-DO).}\label{fig:as-fair-do}
\end{figure*}

%% file: section/technologies.tex
\FILE{section-technologies.tex}

\subsection{Enabling Technology Concepts and Terminology}

A number of technologies are enabling us to develop the framework we
describe here, and they provide the cornerstone of our efforts.

\begin{description}

\item[Cloud]
     According to NIST, cloud computing ``is a model for enabling
     ubiquitous, convenient, on-demand network access to a shared pool
     of configurable computing resources (e.g., networks, servers,
     storage, applications, and services) that can be rapidly
     provisioned and released with minimal management effort or
     service provider interaction.'' The model is composed of five
     characteristics addressing together {\em on-demand self-service},
     {\em broad network access}, {\em resource pooling}, {\em rapid
     elasticity}, {\em measured services}, {\em software as a
     service}, {\em platform as a service}, {\em infrastructure as a
     service}, and {\em private, public, hybrid cloud},

\item[Hyperscale cloud compute centers]
     provide compute centers that are scaled based on increased demand
     by the user. Hence users have seemingly access to resources they
     require. Such centers continuously update server, network, power
     and other resources to meet demand while offering services for
     rent.

\item[Leadership class computing facilities] 
     In the US and also worldwide \cite{www-top500} government
     agencies have worked towards making Leadership Class Computing
     Facility (LCCF) is available to the research community. Such
     facilities provide large-scale computing and data resources. In
     the US, they are funded by DOE, NSF, and other
     agencies. While the first exascale computer was delivered at
     ORNL \cite{www-top500} other systems will become online over the
     next three years. Together the LCCF will comprise an {\em ecosystem
     for very large-scale computing in support of promoting progress
     in science.''} They are expected to deliver a significant boost
     in the capable computing power, addressing some of the grand
     challenges. As such systems are complex and researchers desire
     ease of access, a service model provides one way of accessing
     them.

\item[Representational state transfer (REST)]
     is a software architectural style in support of the design and 
     design and development of services exposed to the  World Wide
     Web. REST defines a set of 
     constraints for how the services behave while focusing on 
     scalability of interactions between components, uniform
     interfaces, independent deployment of components, and the
     creation of a layered architecture to facilitate caching
     components to reduce user-perceived latency, enforce security,
     and encapsulate legacy systems \cite{www-rest}.

\item[Microservices]
     are an architecture style ``to describe a particular way of
     designing software applications as suites of independently
     deployable services. While there is no precise definition of this
     architectural style, there are certain common characteristics
     around organization around business capability, automated
     deployment, intelligence in the endpoints, and decentralized
     control of languages and data. \cite{www-microservices}''

\item[Analytics as a service]
     provides access to subscription-based data analytics software
     through the cloud. Analytics services may include the
     sophisticated combination of services internally used to provide
     customized offerings to the users. The range of resource
     requirements, including the time needed to obtain an answer, could
     vary widely. For long-running analytics efforts, asynchronous
     services may be offered, allowing to pick up the result of an
     analytics task at a later time.

\item[Data analytics as a service]
     With increased data demands, it is important to integrate the data
     storage needs to access data needed by an analytics service. In
     the case of significant data needs, moving the data
     to a new service is often impractical. Hence the analytics is often conducted as part
     of add-on services running close to the data. This is often referred to as "bringing the calculation to the data.''

\item[Machine and deep learning].
     Machine learning enables analysis of data by  ``learning''
     from the data. Deep learning [DL] is a subdiscipline of machine
     learning. DL introduces sophisticated toolkits and frameworks which use multi-layered neural networks that enable non-linear
     transformations on the data. The network is first trained on
     a subset of data aka input data; and then new data can be fed to the trained model, to
     obtain an output such as a prediction. DL models require extensive data and
     extensive training to be accurate. Another challenge is finding a good model and good hyper parameters to address a particular
     problem.


\end{description}

%% file: section/usecases.tex
\FILE{section-usecases.tex}

\section{Use Cases for Analytics Services}
\label{sec:usecases}

Our work is motivated by a number of use cases. The use cases were
contributed by community members who are experts in their fields.

The use cases include security; numerical weather prediction; 
Heating, Ventilation, and Air Conditioning (HVAC) optimization; and earthquake prediction.

Each use case includes a high-level explanation about the problem they address and 
explicitly comments on requirements needed for an analytics service targeting the individual use case.
We summarize the major requirements for all use cases in Table~\ref{tab:summary-as-requirements}

\input{usecase/table.tex}

%% file: usecase/table.tex
\begin{table}[htb]

\newcommand{\YES}{\checkmark}
\newcommand{\NO}{--}

\caption{Summary of selected AS requirements}
\label{tab:summary-as-requirements}
\begin{tiny}
\begin{tabular}{|l|lll|l|}
\hline
\bf \cellcolor{blue!25} Attributes  & \bf \cellcolor{blue!25} HVAC & \bf \cellcolor{blue!25} Security & \cellcolor{blue!25} \bf Earthquake
& \cellcolor{blue!25} \bf Requirements \\
\hline
\hline
\multicolumn{5}{l}{\cellcolor{blue!10} \it Compute} \\
\hline
Cloud  &  \YES &  \YES &  \YES &  \YES \\
Hybrid & \YES & \YES &  \YES &  \YES\\
Multi-Cloud & \YES & \YES & \YES &  \YES\\
Sensors & \YES & \YES  &  \YES &  \YES\\
LCCF & * & \NO &  \YES &  \YES\\
Microservices & * & \YES &  * &  \YES\\
Vendor neutral & \YES & \YES &  \YES &  \YES\\
REST & \YES & \YES &  \YES &  \YES\\
\hline
\multicolumn{5}{l}{\cellcolor{blue!10} \it Workflow} \\
\hline
Catalog & \YES & \YES &  \YES &  \YES\\
Cooperation & \YES &  \YES &  \YES &  \YES\\
Competition & * & \YES &  \YES &  \YES\\
Orchestrator & * & \YES &  \YES &  \YES\\
\hline
\multicolumn{5}{l}{\cellcolor{blue!10} \it Calculation} \\
\hline
Modeling & \YES & \YES &  \YES &  \YES\\
Calculation & \YES & \YES &  \YES &  \YES\\ 
Deep Learning & \YES  & \YES &  \YES &  \YES\\
Visualization & \YES & \YES &  \YES &  \YES\\
\hline
\multicolumn{5}{l}{\cellcolor{blue!10} \it Security} \\
\hline
Communication & \YES & \YES &  \NO &  \YES\\
Data & \YES & \YES &  \NO  &  \YES\\
Access & \YES & \YES &  \YES &  \YES\\
Usage policies & \YES & \YES &  \NO &  \YES \\
\hline
\multicolumn{5}{l}{\cellcolor{white} * = Future} 
\end{tabular}
\end{tiny}
\end {table}

%% file: usecase/hvac.tex
\subsection{Use Case: HVAC Recommendation}

\paragraph*{Background.}
Heating, ventilation and air conditioning (HVAC) systems control air temperatures inside residential, commercial, and industrial  buildings. 
Desired room temperatures can be predetermined/fixed by a technician, or regulated at any time by the user, through the use of a local thermostat, where the user chooses temperature setpoints. This manner of  HVAC control only considered user preference inputs at the moment the thermostat is programmed. Modern thermostats are capable of applying machine learning to leverage historical data on user preferences, and existing periodic variances in energy and  electricity prices. To improve the HVAC system efficiency over time, introduction of other criterion and inputs into temperature setpoint decision making can reduce costs for consumers and in a scenario involving a million systems, present a potential for electricity service providers to better account for fluctuating consumer demand; and balance load. Here we are also including weather forecasts to the list of external inputs or integrating weather research forecast WRF and we develop an algorithm that can calculate the optimal schedule of HVAC control. 

\begin{figure}[htb]
\centering\includegraphics[width=1.0\columnwidth]{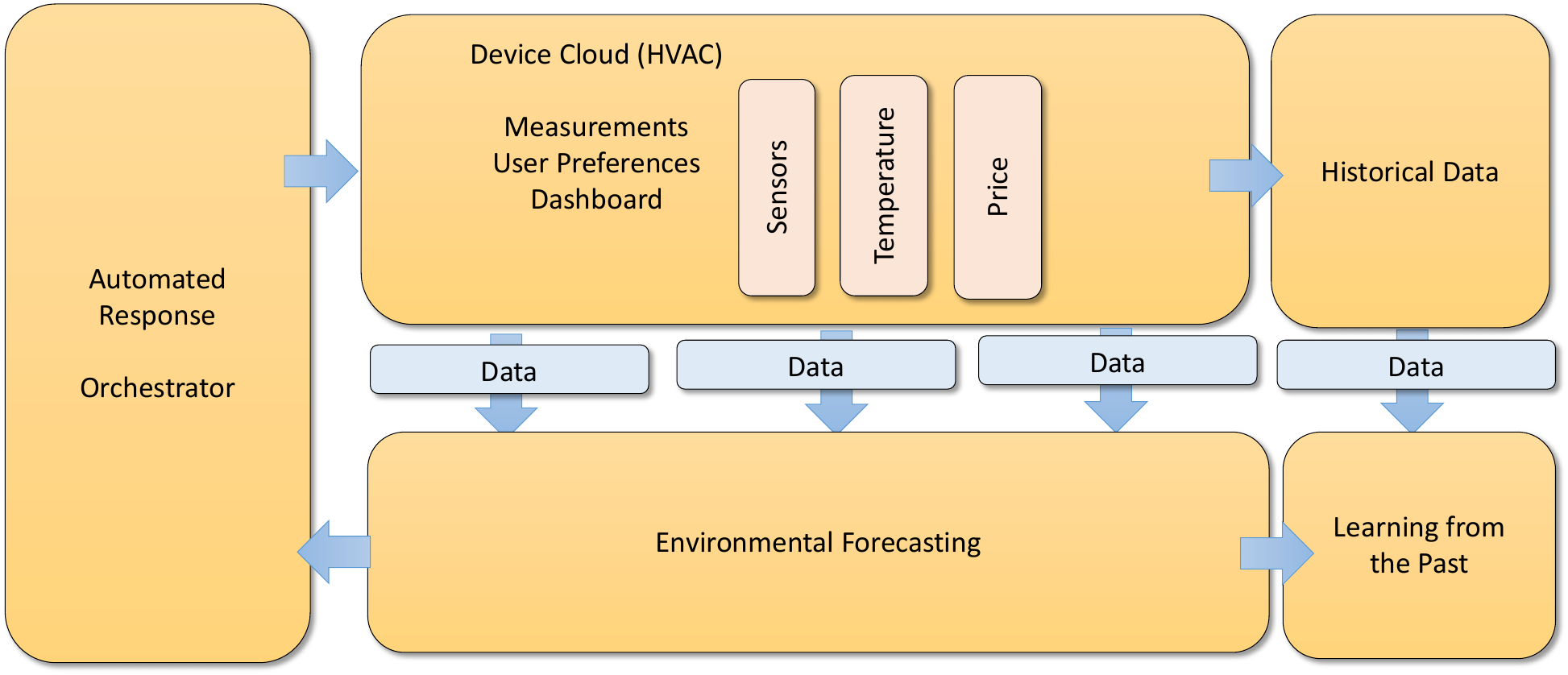}
\caption{General view of HVAC-to-Cloud system}
\label{fig:hvac_general}
\end{figure}

The algorithm deployment is on the cloud as a service, and every time it makes a new setpoint decision, the command is sent to the local thermostat (see Figure \ref{fig:hvac_general}). Various service frequencies were used to establish the proper pipeline for this application use case. This work aims to analyze the scalability and adaptability of such service scenarios and identify best practices that promote reusable implementations to support aspects of similar use cases addressed by them.

\paragraph*{System model.}
User comfort level is defined as the allowed temperature range in the house, the range can remain constant during the entire 24-hour cycle, or it can vary based on the time of the day or the day of the week. Since the temperature in the house is affected by the outdoor temperature our goal is to keep the internal temperature within the desired user comfort level. 
Energy prices fluctuate throughout the day, driven by market dynamics. We describe the pricing function as a sequence of values $Price = {P_{t_1}, P_{t_2}, . . . , P_{t_m}}$ where $P_{t_i}$ corresponds to the price at time slot ${t_i}$. At each time slot, the service provider determines the electricity price and charges the customer at the given rate. We determine the user cost based on the price and energy consumption during the time slot. At time $t_{k-1}$, we observe the set of all variables of the system which includes information about the outdoor temperature, the indoor temperature, and the pricing. Based on all this information, the algorithm makes a dynamic decision to set the HVAC set point. This represents the HVAC state in the next time interval $t_k$. After executing the HVAC setpoint action, we observe the system during time step $t_k$. These steps are executed in periodic intervals, as presented in Figure \ref{fig:hvac_general} and the setpoint recommendation process requires an advanced machine learning algorithm (e.g, reinforcement learning) \cite{kotevska2020rl}. 

We organized this functionality into three parts: Data input (DI), Functionality parts (FP), and RL agent (RLA). DI collects the data needed for the calculation, such as the following: 
\begin{itemize}
    \item Temperature -- Collects current weather temperature.
    \item Price -- Collects current electricity price.
    \item Historical data -- Extract needed data fields and packs it into an intermediate file format. Input data from the output of temperature and price.
\end{itemize}

FP prepares the inputs for RLA, they are the following:
\begin{itemize}
    \item Learning from the past -- Learns from previous user and system activities.
    \item Environment forecasting -- Incorporates future price and temperature forecasting for timestamp ${t_{i+1}}$.
    \item Reward -- Generates reward based on the current temperature, price, and user input.
    \item User preferences -- Creates rules based on user temperature preferences.
\end{itemize}

RLA model estimates the next set-point ${t_{i+1}}$ using the inputs from FP, such as: 
\begin{itemize}
    \item RL agent --  Interpolates the output from DI, FP, reward, and user preferences and generates action recommendations for the temperature.
\end{itemize}

Figure \ref{fig:hvac_general} shows the general modeling system flow chart. 







Reward is a function of temperature violation and cost. This function is specific to the algorithm that was used in this case reinforcement learning (RL). If the temperature is above the desired setpoint and energy cost is high the reward is negative. While if the temperature is within the setpoint range and the energy price is low the reward is positive. This allows the RL-based method to learn the optimal behavior through continuous interaction with a building environment and without referring to any prior model knowledge. 

The results are presented on the home device interface and historic results are presented on the dashboard using the cloud-as-a-service option.

This setup is a general setup that is applicable to one-zone models. Results have shown that RL model has high generalization and adaptability to unseen environments, which indicates its practicability for real-world implementation \cite{du2021intelligent}. The same functionality can be used with different building models and retail price models \cite{du2021intelligent}. 



This use case has the implicit requirements needing the following aspects to be addressed by the framework we develop.

\begin{enumerate}

\item{\bf AS vendor neutral cloud and computer service integration.} 

\begin{enumerate}
  \item {\bf AS in cloud.} The Analytics Services are on the cloud, and the user has Vendor Neural Access Interface to monitor the current and past behavior. The users can also manage their preferred HVAC behavior using the provided interface by the command line.
  \item {\bf AS in LCCF.} This feature does not apply to the current use case. However, an extension could be addressing larger scales of the application while integrating analysis on a much larger scale.
  \item {\bf AS in microservices.} This feature does not apply to this use case. We anticipate that future services may use microservices in order to increase portability.
\end{enumerate}

\item{\bf AS architecture.} 

\begin{enumerate}
  \item{\bf AS vendor neutral interfaces.} The user has access to two interfaces: a) a web-based dashboard to monitor the current and past HVAC status changes and indoor temperature; b) an HVAC-embedded home interface where the user can check and change the desired indoor temperature.
  \item{\bf AS REST.} The HVAC home interface accepts user preferences and sends them to the cloud using REST services. 
  \item{\bf AS layers such as interface, service layer, and provider layer.} The vendor uses Privacy Cloud for the calculations and data storage, Analytics Services for decision-making analysis, Cooperation Service for external data inputs, and command line for the user to send requests and make any service adjustments.
\end{enumerate}

\item{\bf AS workflow.} 

\begin{enumerate}
  \item{\bf AS catalog and registry.} For this use case, the catalog of services only supports a few operations they are i) temperature set-point adjustment based on outdoor temperature, price, user schedule, and preferences; ii) historical overview; iii) receiving user input for set-point adjustment.
  \item{\bf AS cooperation.} The cooperation is with external services used for decision-making analysis (e.g., weather forecast, utility energy price).
  \item{\bf Competition.} This feature does not apply to this use case. In the future alternative analysis algorithms could be integrated that allow collaborative results but also competition.
  \item{\bf AS orchestrator.} This feature does not apply to this use case. However, an extension could be envisioned that allows the execution of the analysis on different geographically distributed resources to circumvent outages of the center due to unforeseen challenges, such as extreme weather events or power outages.
\end{enumerate}

\item{\bf AS calculation.}

\begin{enumerate}
  \item{\bf AS with DL.} In this case is used DL ( e.g., reinforcement learning (RL)) for the decision-making analysis.
  \item{\bf AS data analytics.} In this use case, data analytics provides visualization and statistics for the past behavior and decision-making recommendations for HVAC setpoint settings.
\end{enumerate}

\item{\bf AS security.} The data transfer to and from the cloud is over a secure protocol.

\end{enumerate}

%% file: usecase/security.tex
\subsection{Use Case: Continuous Monitoring for Enterprise Security}

\paragraph*{Background.}
Most modern-day big corporations have a hybrid multi-cloud architecture 
with points of presence on-premise and multiple cloud vendors. 
Secure-by-design is a key architectural consideration for these 
enterprises as a safeguard against organized cybercrime activities 
such as ransomware and sensitive data exfiltration. Continuous monitoring 
is one of the key elements for secure-by-design whereby critical 
assets and network communication are continuously monitored for signs 
of suspicious behavior and any threats identified are responded to in an automated way.

\begin{figure}[htb]
\centering\includegraphics[width=0.8\columnwidth]{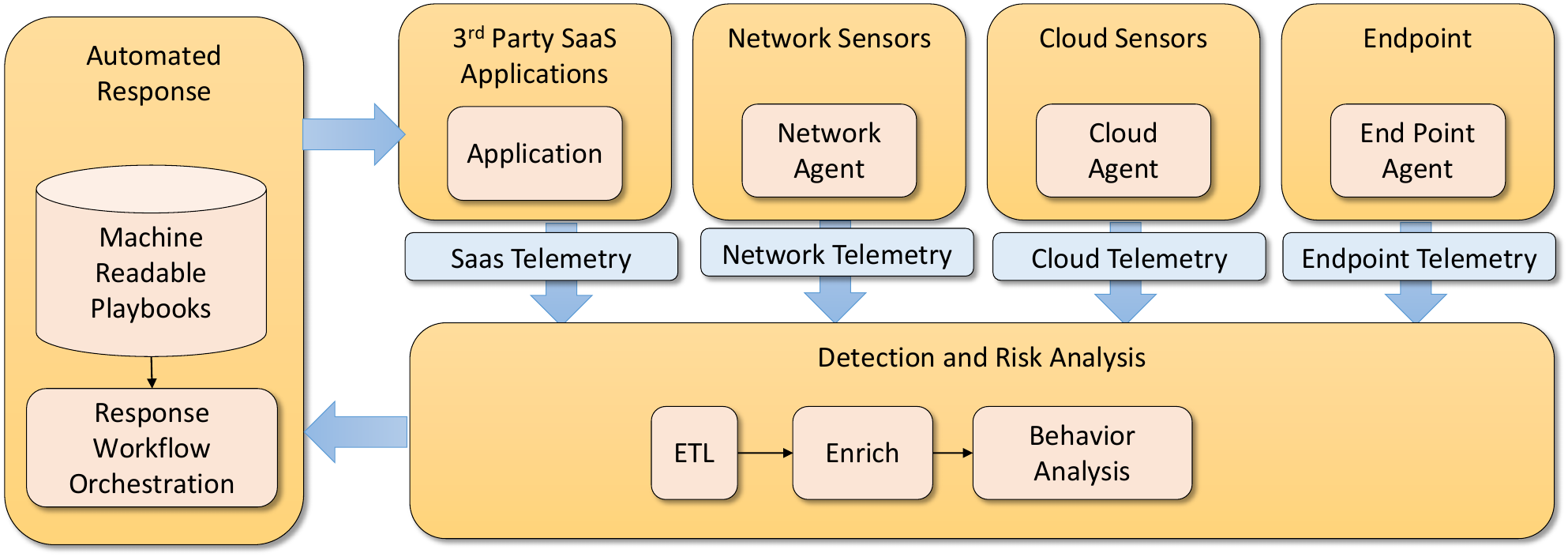}

\caption{Continuous Monitoring and Response}
\label{fig:sec-general}
\end{figure}

Figure \ref{fig:sec-general} shows a simplified version of a typical 
Continuous Monitoring Pipeline which usually consists of distributed services for the following functions:

\begin{enumerate}

\item{\bf Telemetry Collection.} Telemetry is typically collected from an endpoint, 
network, cloud, and 3$^{rd}$-party Software as a Service (SaaS) applications by endpoint, network, and cloud agents, 
firewalls, Web Application Firewall (WAF), email security, and vulnerability scanners. Telemetry usually consists 
of operating system (OS) and application audit logs, vulnerability scan results, software Bill Of Materials (SBOMs), east-west and 
north-south network packets and flow logs, cloud policies etc.

\item{\bf Detection and Risk Analysis.} The Detection and Risk Analysis pipeline 
comprises  extract, transform, and load (ETL) for data normalization, enrichment on normalized data by correlating 
with threat intel from inhouse and 3$^{rd}$ party services, and asset profiling and
behavioural analysis for misbehaviour detection which is done using rule-based 
and machine-learning-based approaches.

\item{\bf Automated Response.} Automated Response consists of security response 
playbooks and workflows for alerting, breach containment, mitigation, remediation 
and asset recovery.

\end{enumerate}

This use case has the implicit requirements needing the following
aspects to be addressed by the framework we develop.

\begin{enumerate}

\item{\bf AS vendor-neutral cloud and computer service integration.}

  \begin{enumerate}
  
  \item {\bf AS in cloud.} The Analytics Service for this use case is typically hybrid with points of presence on-premise where analytics modules can run on the sensors and on cloud. On-premise sensors can have support for local storage and running light-weight ETL and machine learning inference, whereas cloud can have more heavy-weight compute and machine learning model training and inference support.
  
  \item {\bf AS in LCCF.} This feature does not apply to the current use case.
  
  \item {\bf AS in microservices.} Microservices approach is highly applicable to the design of continuous monitoring pipeline where each of the subfunctions for telemetry collection, ETL, enrichment, analytics and automated response can consist of several microservices that could either be developed in-house or leverage 3rd party SaaS.
  
  \end{enumerate}

\item{\bf AS architecture.}

  \begin{enumerate}
  
  \item{\bf AS vendor neutral interfaces.} Vendor-neutral interfaces would be ideal for different functions of the Continuous Monitoring Pipeline. However, at this time there are no standardized interfaces for technologies related to continuous monitoring. Each vendor typically has its own proprietary set of interfaces and data models and it is up to the consumer of the data to normalize that data.
  
  \item{\bf AS REST.} REST is typically used for human and M2M interfaces at different levels in the Continuous Monitoring pipeline. On-premise and cloud agents for telemetry collection support both data push and pull via REST interface. Data Enrichment services can be in-house and 3rd party SaaS which typically have REST interfaces for pulling reference datasets. Analytics and Automated Response pipelines typically have REST interfaces for pipeline configuration and adding or updating content.
  
  \item{\bf AS layers such as interface, service layer, and provider layer.} Interface layer is required for data access for experimentation and usecase development, for configuring and adding new functions and content to telemetry collection, analytics, and response functions. The service layer is required to register in-house and 3rd party microservices to support the continuous monitoring pipeline and to register analytics workflows. Provider layer is required to schedule services and workflows on-premise and on cloud.
  \end{enumerate}

\item{\bf AS workflow.}

  \begin{enumerate}
  
  \item{\bf AS catalog and registry.} Continuous Monitoring will comprise several microservices for telemetry collection, ETL, enrichment, analytics, and automated response. These services can be in-house or 3rd party SaaS services running remotely and accessible via APIs. The catalog and registry functionality will be required to catalog and register these services and have required configuration for service setup and service access.
  
  \item{\bf AS cooperation.} There can be intra and inter-function collaboration between the services that are part of the Continuous Monitoring Pipeline. For instance, endpoint, network and cloud agents belong to the telemetry collection function and can have intra-function collaboration where they can cooperate with one another on what telemetry to collect on the fly based on contextual and situational awareness. There is also inter-function collaboration for instance between automated response pipeline and endpoint, network and cloud agents for breach containment, mitigation and asset recovery.
  
  \item{\bf AS competition.} There can also be competing services and rules can be defined on how to orchestrate between these competing services. For instance, for instance, the Continuous Monitoring Pipeline can have multiple 3rd party and in-house services for threat intelligence enrichment and rulesets can be defined on how to orchestrate between these services to streamline cost, speed, and resources.
  
  \item{\bf AS orchestrator.} API-based workflow definition and orchestration capabilities are required for the Continuous Monitoring Pipeline to orchestrate cooperating functions and cooperating and competing microservices that are part of those functions.
  
  \end{enumerate}

\item{\bf AS calculation.}

  \begin{enumerate}
  
  \item{\bf AS with DL.} Deep Learning is leveraged in the analytics stack for behavior profiling and misbehavior detection and knowledge graph analysis (e.g., graph neural networks and neuro symbolic analysis).
  
  \item{\bf AS data analytics.} Data analytics is leveraged for data visualization and model quality and drift detection, and visualizing and report generation on security outcomes and alerts. 
  
  \end{enumerate}

\item{\bf AS security.} AS Security is required for securing data at rest and during transit, for securely storing service access credentials, and for enforcing resource access and usage policies. 

\end{enumerate}

%% file: usecase/earthquake.tex
\subsection{Use Case: Earthquake Prediction Benchmarks}

Earthquake prediction of their occurrence and likely-hood is a complex problem. 
It is difficult because the details of
the underground plates and the friction laws between them are not
known and furthermore, the earthquake causes phase transitions between
plates which makes for unpredictable movement in addition to 
predicting earthquake energy wave movements. 
Since both of these problems
are so complex with many hidden variables and the equations governing
the phenomenon is unknown or incomplete, deep learning can be use to
learn the various hidden patterns in the data which the model can then
use to forecast earthquakes into the future.
The application of models and hyper-parameter sets is an open challenge requiring a large amount of computational power and the investigation of several promising models.

\begin{figure}[htb]
\centering\includegraphics[width=1.0\columnwidth]{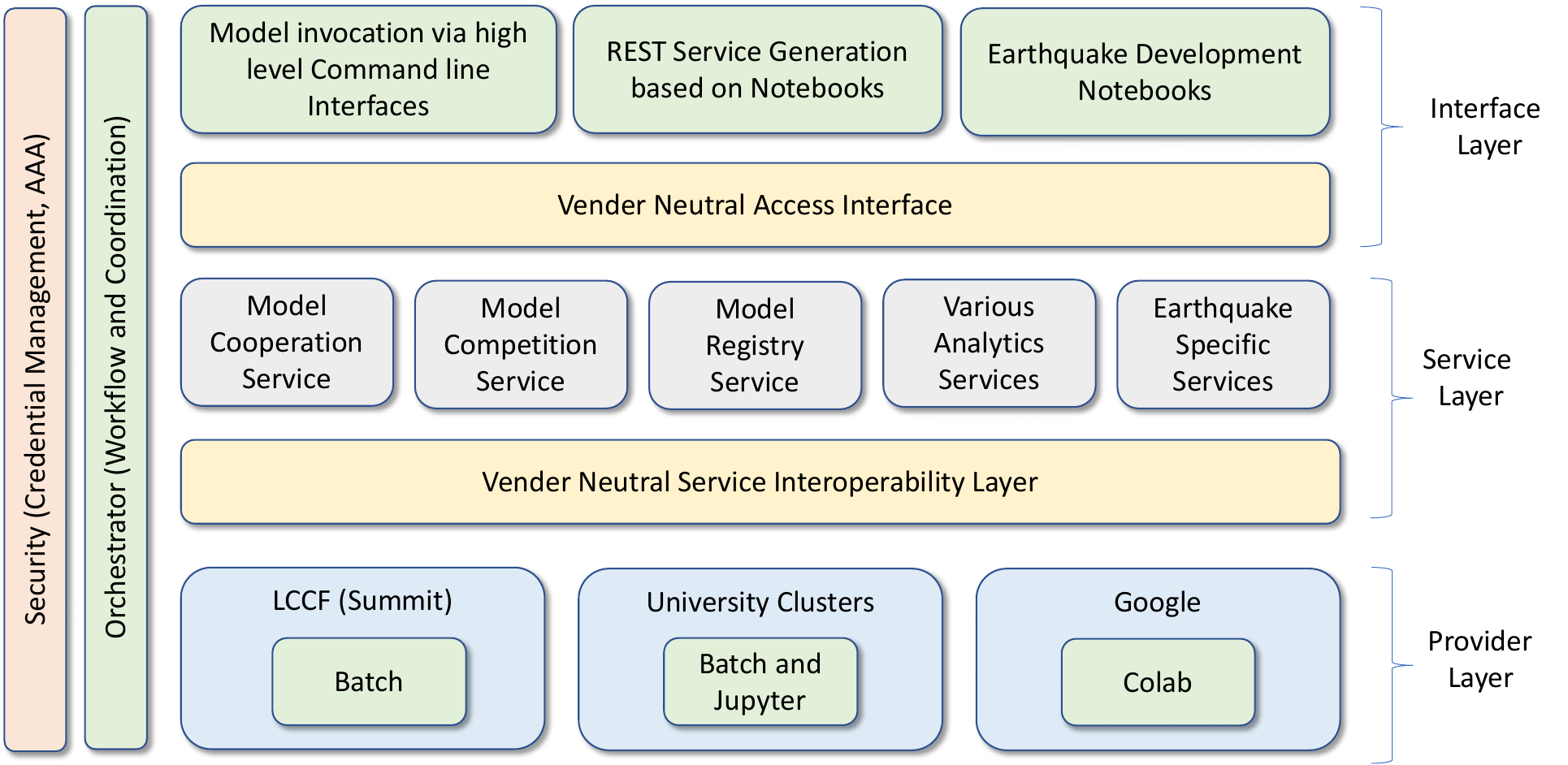}
\caption{General view of Earthquake Prediction Benchmark system}
\label{fig:eq-general}
\end{figure}

As part of this effort a benchmark activity has been started within the MLCommons Science working Group \cite{las-2023-mlcommons-edu-eq}. The usecase deriving from this effort motivates our efforts. Figure \ref{fig:eq-general} shows the architecture of our current efforts which can be generalized to be applicable to many scientific analysis problems.

This use case has the implicit requirements needing the following
aspects to be addressed by the framework we develop.

\begin{enumerate}

\item{\bf AS vendor-neutral cloud and computer service integration.}

  \begin{enumerate}
  
  \item {\bf AS in cloud.} The analysis of this use case requires a significant amount of computing time as well as specialized hardware such as the utilization of GPUs. It is beneficial that the analytics services developed can run on hybrid clouds to leverage best availability and cost. Development can be supported by on-premise resources. 
  
  \item {\bf AS in LCCF.} The availability of large HPC resources if of advantage as the calculation and hardware needs for this problem exceed those of a desktop and can benefit from running many of the models in parallel to select the best once producing a minimal scientific accuracy measurement.
  
  \item {\bf AS in microservices.} Although microservices can be used in this effort it has not been much applied due to the greater need of accessing HPC resources.
  
  \end{enumerate}

\item{\bf AS architecture.}

  \begin{enumerate}
  
  \item{\bf AS vendor neutral interfaces.} Vendor-neutral interfaces are beneficial as it would allow analytics services to be integrated in a plug-in fashion.
  
  \item{\bf AS REST.} REST services would be beneficial as the highly specialized analytics services and resources can be accessed through language and implementation-neutral interfaces. This would allow the access of services by non-experts, but also the integration of services that are developed by communities.

  \item{\bf AS layers such as interface, service layer, and provider layer.} As The interface layers provide the necessary abstractions to benefit from efforts across implementers specializing in various layers such as scientists, interface designers, cloud providers, and HPC providers. 

\end{enumerate}

\item{\bf AS workflow.}

  \begin{enumerate}
  
  \item{\bf AS catalog and registry.} It is beneficial to register the 
  various models and their inputs and outputs as well as performance needs and results. This way components can be chosen based on the implicit analytics performance characteristics. For example, the model of an earthquake near a volcano could be very different from a model close to the movement of a tectonic plate.

  \item{\bf AS cooperation.} The need of cooperating services and models can enhance the accuracy of the resulting new model calculation. 
  
  \item{\bf AS competition.} The integration of models found in the future and a dynamic reanalysis of the analytics functions performed earlier in case a better model is found.
 
  \item{\bf AS orchestrator.} API-based workflow definition and orchestration capabilities are required for coordinating the complex environments and analytics functions.
  
  \end{enumerate}

\item{\bf AS calculation.}

  \begin{enumerate}
  
  \item{\bf AS with DL.} Deep Learning is leveraged due to the complexity of the problem and the inaccessibility of details of the earth mantle.
  
  \item{\bf AS data analytics.} Data analytics is leveraged for generating the forecasts on a spatial and time-dependent level.
  
  \end{enumerate}

\item{\bf AS security.} AS Security is used to control access to the various resources. As the research is carried out in an Open Source community other security issues are of little concern. 

\item{Data needs}  The data needs can become very large if real-time sensors are integrated. However, smaller data sets are available and are currently used by this application to analyze Earthquakes in California from 1950 till today.

\end{enumerate}



%% file: section/architecture.tex
\FILE{section-architecture.tex}

\section{Architecture}

To support the usecases we briefly described and analyzed their requirements, 
we have developed a general architecture that can benefit most analytics services while leveraging and integrating hybrid and multi-cloud analytics services.

\begin{figure}[htb]
  \begin{center}
    \includegraphics[width=1.0\columnwidth]{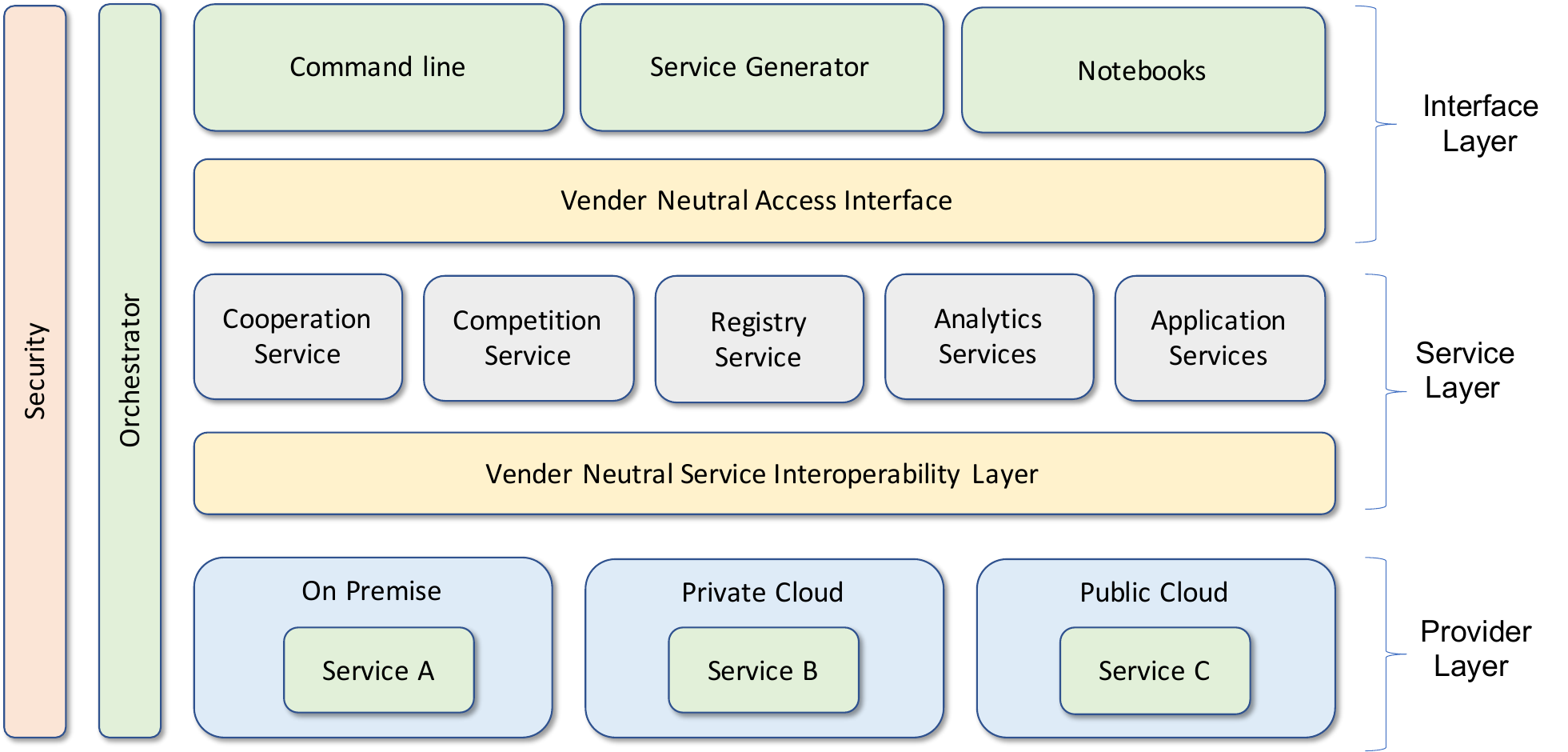}
    \end{center}
  \caption {Architecture of the hybrid multi-analytics service
    framework}
  \label{fig:arch}
\end{figure}

To support our goal to enable the use of hybrid and multi-cloud analytics
services, we are exploring architectural patterns
that are conducive to use cases such as the one we outlined
previously. These patterns are of general use as they can
be applied to other use cases. In our case, we define an
architecture as depicted in Figure~\ref{fig:arch}.

The architecture is organized in layers and contains
multiple components in each layer. We distinguish the interface,
service, and provider layer. Security and an orchestration service
enable the integration of the various components into a coordinated
application pipeline.

\subsection{Interface Layer}

Today's analytics is invoked through a multitude of interfaces, making
it possible to invoke them in different languages, but also high-level
frameworks. This is often achieved by an interface layer using REST
to communicate with the other layers. For our work, we focus on
the command line interface promoting reuse in shells, Jupyter notebooks
showcasing the reuse in an interactive analysis capable framework, and
our previous work to generate REST services (see Section \ref{s:gas}).

\subsubsection{Command Line Interface}

To provide reusability within the DevOps application of data analytics
pipelines, we provide an enhanced command-line interface that
specifically targets hybrid and multi-analytics environments. This is
facilitated by adding command options, shell variables, and
configuration files that can be passed into the commands.
Most importantly, we analyze which specific parameters we
need to make available when investigating cooperative and competitive
services. The parameters can be directly translated to REST service invocations.

\subsubsection{Interactive Notebooks}

Although it is important to provide a command-line interface that can
easily be used to generate computing activities for solving data
analytics tasks, it is even more critical that the framework can be
integrated into interactive steering tasks conducted directly by the
data scientists. For this, we leverage Jupyter notebooks and integrate
service calls to the backend system into the notebooks just as we can
in regular Python programs. The difference here is that Jupyter
provides a rich set of interactive components and widgets that can be
leveraged to prototype interactive services but also parameter
studies. While our
previous work has been integrated into notebooks, the capabilities of
notebooks were not yet fully utilized as the integration was done on
the service level but not on a level where Jupyter was used to enhance
the service pipeline. 

\subsection {Service Layer}

\subsubsection{Generalized Analytics Services}
\label{s:gas}

While we have focused previously on the automated generation of
services using REST, we need to consider the other aspects of this
architecture that are needed to support the data analyst. We have
shown that it is possible to create rest services from Python
functions and classes while augmenting them with helper services such
as data uploads. The development of such services is out of the
capability of many data scientists as they focus on developing
transformative data analysis functions and not on the infrastructure
service generation. Our work is lowering the bar for such
implementation and allowing even data analysts to generate reusable
REST services \cite{las21openapi}. The effort to learn how to
create vendor-independent and computer language-independent services
has been reduced from months to days. We can leverage this effort to
generate application-specific services quickly. The services generated
are integrated into a user-managed service registry. We leverage
this service and enabled its exposure via REST through FastAPI as part
of the generator.

\subsubsection{Hybrid Multi-Analytics Service Registry}


While we have previously developed a simplified generalized service
registry, we are exploring significant extensions to integrate (a)
container images, (b) container services, and (c) analytics 
services offered by service providers. This registry is specially
designed to support cooperating and competing for analytics services.
We intend to add the ability to leverage existing repositories, such as
GitHub and DockerHub to register suitable analytics services as source
code. We will then also add features to provide endpoints to
instantiated services so they can be advertised to a large group of
users. A neutral vendor specification is used as part of the
registry. Such a registry can be hosted by a user or an
organization. We will identify if it is possible to leverage GitHub
for hosting such a registry. This will require a special set of tools
and programs to keep the registry up to date. A user can then
integrate such a registry into their analytics pipeline. New analytics
service specifications can then be either integrated through direct
specifications added to the hosted registry or through the use of
GitHub submodules. Using submodules offers the ability to keep up to
date with analytics services developed by others and allows updates
through automated DevOps-controlled pull requests. This registry
technology would completely replace our earlier registry work if
successful. It would also allow the integration of private services as
private analytics services can be integrated while using private
submodules hosted in private repositories. Hence, the details of such
modules are not exposed. As GitHub also supports GraphQL we will
explore using GraphQL as a mechanism for the specification of Registry
entries. To increase privacy, git can also be hosted on-premise.

\subsection{Application Services}

An application may require the availability of very specific
application-oriented analytics services. Our architecture allows us to
integrate them while reusing the same vendor-neutral specification
format. This includes not only cloud services but also the
integration of analytics services that rely on on-premise
infrastructure. An example would be access to a supercomputer in the
TOP500 list that is used to conduct a complex data analytics task
reusing GPUs to conduct deep learning for COVID-19 analysis. This
results in two specifications. A general specification that can be
reused on other similar on-premise computers, and a second that is
specifically targeted to the targeted on-premise infrastructure. This
could include the integration of hosted data services or specific
network capabilities.

\subsection{Analytics Services}

Data analysts are developing analytics functions on a regular
basis. As we can use our service generator to transform them into
analytics services, we will be able to create and register them into
our registry. We will expand upon our available services but focus on
services that explicitly address multi-cloud and hybrid service use
cases.

A good example may be natural language processing to analyze a text
with either a local service, a loud hosted service by different
vendors. Here, based on input parameters, we create an overarching
language analytics service that chooses the various services with the
the help of service level requirements and agreements.

\subsection{Cooperation and Competition Services}

As previously indicated, we already have identified two special use
cases of data analytics services that leverage hybrid and
multi-analytics 
services. This is the specification of services that
employ:

\paragraph{Cooperation.} Cooperating services are services that use one or
more services from hybrid multi-analytics services. They are
cooperating together to address the solution to a formidable data
analytics problem. Thus the resources form a "team" of services that
work together. This includes the integration of specialized services
that may not be unique to a particular provider. Still, it also could
mean that computational analytics processes could be performed in
parallel, and results could be gathered to accelerate the analytics
task. A parameter study is a very good example of one kind of
cooperating services

\paragraph{Competition.} Here, the available hybrid multi-analytics services
directly compete with each other. This can be done by direct selection
of services that are more suitable than others. This selection could
be based on resource requirements such as time, availability, cost,
and features. However, the framework could take "observations" and
propose automated conclusions about which services to choose from. A
possibility would even be to integrate deep-learning strategies into
the selection process.

\subsection{Provider Layer}

An integral part of the proposal is identifying how we can leverage
services from multiple providers, including on-premise services. We
have shown in our previous work that we can specify vendor-neutral
specifications to access, for example, virtual machines. We will expand
this concept while using the concept of containers. However, we also
need to identify services that are offered by
multiple vendors, such as language processing services. Although they
can be directly accessed via vendor-specific interfaces. It will
be important to identify if they can be generalized so the users
can benefit from a uniform vendor-neutral service interoperability
layer. We will identify a usability example to explore the
possibilities of this approach.

\subsection{Crosscutting Services}

We have two crosscutting services. One is the {\em orchestrator} that
allows the specification of service pipelines to combine the various
services that are needed for application implementation. The other
is a {\em security} service that will enable us to access the various services
through the required authentication and authorization mechanisms. The
latter we have demonstrated in {\em cloudmesh} where users can manage their
own access to a multi-cloud environment ta access their activated
analytics services. We will leverage from that effort but  also leverage from
open-source solutions that can be embedded in our vendor-neutral
service specification, such as basic and OAUTH security. In general, we
abstract the security calls to be callouts to the appropriate
authentication mechanisms.

%% file: section/security.tex
\FILE{section-security.tex}

\subsection{Security}
\label{sec:security}

Analytics services have a variety of security needs. This includes
authentication, authorization and audit (Section \ref{sec:aaa}). Data
and use may also need to be secured and privacy concerns need to be
addressed (Section \ref{sec:privacy}).

\subsubsection{Authentication, Authorization and Audit (AAA)}\label{sec:aaa}

Authentication and Authorization functions would be needed across the
Analytics Service Framework to authenticate and authorize
human-to-machine (H2M) and machine-to-machine (M2M) communication
between the interface layer and the service layer and between the
service layer and the provider layer.

Communication to the services in the Service Layer would be via
RESTful APIs. For authenticating to services in the Service Layer,
Single-Sign on (SSO) using token-based access control schemes like
OAuth, SAML, or Kerberos will be used. Most Web Service frameworks
support these SSO mechanisms out-of-the-box. Once the client is
authenticated, authorization checks are done to ensure the client is
authorized for the requested CRUD operation on the REST resource.

Authentication and authorization for machine-to-machine (M2M) access
between the service layer and Private and Public Cloud in the provider
layer will be done using Cloud's Role-based Access Control (RBAC) and
Attribute-based Access Control (ABAC) functions and Identity and
Access Management (IAM) policies. Service accounts are provisioned in
the cloud environment for service clients, and these service accounts
are given appropriate roles based on the Principles of Least Privilege
(PoLP).

Client actions such as access to resources, changes to configuration, and changes to roles and access policies should be continuously
monitored and audited. Most web service frameworks have basic audit
and logging functions that can create audit trails on files,
databases, or remote logging services. Audit on cloud providers can be
done by enabling cloud audit trails and cloud alerts.

\subsubsection{Privacy}\label{sec:privacy}

The Analytics Service Framework will have appropriate controls in
place to ensure data privacy. The framework will support encryption
of data at rest and in transit, both intra-layer and inter-layer, to
safeguard against unwanted access to data. Any data that is copied
over for processing purposes by the framework will be deleted once the
processing is done. The framework will also allow administrators to
delete inactive users and service accounts and revoke
accesses. Sensitive attributes in the data will be masked from users
and services based on roles and policies. The framework will also
audit and log any access to data which can then be monitored in a
Security Incident Event Management System (SIEM) for unwanted or
abnormal access.

There are some other privacy controls that the broader ecosystem
should have which would be out of the scope for the Analytics Service
Framework because they don't belong there. Some of these include
having a data inventory to map data storage, classification of all
data and governance processes in place to ensure data literacy and
manage data lifecycle within an organization.

\subsubsection{Federation}

As some services in the analytics framework may be distributed across
a variety of services owned by different providers it is beneficial to
allow the integration of a service federation. NIST has spearheaded an
extensive document \cite{nist-Lee2020} addressing such federation that
could be leveraged and federated services could be developed based on
is.

%% file: section/defining.tex
\FILE{section-defining.tex}

\subsection{Defining and Finding Reusable Analytics Services}
\label{sec:defining}

Defining and Finding Reusable Analytics Services is an important
aspect of the usability of services. To ensure findability we need a
Service catalog and a service registry.

%% file: section/catalog.tex
\FILE{section-catalog.tex}

\subsubsection{Analytics Service Catalogue}
\label{sec:catalog}

\paragraph*{Motivation.}
Cloud providers offer a considerable set of analytics services to
their customers. There are many analytics services available, and a user
needs to be able to quickly obtain an overview of such available
services. This helps identify further actions to evaluate
them and identify if further investigation is justified. The catalog
contains enough details to locate the service and evaluate its
usefulness. However, it may not provide technical details 
captured by a service registry instead.

\paragraph*{Access Requirements.}
The catalog may be public or may be restricted while authorized
entities may access it. As analytics services may evolve. Hence,
time-dependent versioned descriptions of the services must be able to
be included. An organizational entity may manage its own
catalogs. It is desirable to have the catalogs be uniform so that
they can be combined into a larger catalog combining entries of
multiple organizations.

\paragraph*{Federation.}
The offerings are typically limited to a particular vendor. Users can
benefit from a federated service catalog to search and explore for
needed services by the user. In contrast to a registry, a catalog may
not include all technical details but could, in contrast, include
services that lack such details and thus can be the basis of an
exploratory process. A Federated analytics service repository is
could be hosted on GitHub. The catalog contains the
following attributes, many of which are also used in an analytics
service registry.

The catalog is organized as a list of entries, where each entry
contains a number of attributes. These attributes may be required or
optional. We list in Table \ref{tab:reg} in the column Catalog.

%% file: section/registry.tex
\FILE{section-registry.tex}

\subsubsection{Analytics Service Registry}
\label{sec:registry}

\paragraph*{Motivation.} 

The goal of a federated analytics service registry is to establish
federated registries to locate and consume analytics services with
persistent identifiers across organizations.

A service registry can serve as a public, private, or federated
registry. The first two properties define whether the registry is public or
private. In the case of a private registry, proper security measures need
to be taken into account to govern access. Our framework does not make
any recommendations about the security framework chosen and it is up
to the implementer to specify it. In the case of a federated registry,
more than one registry can be joined, to provide the user the
impression of a single registry.

Within the analytics services, we distinguish two classes. The first
class are instantiated (running) services that are offered by a
service provider and allow direct reuse. The second class are library
providers that distribute analytics activity not as an instantiated
service, but as a source code library which can be deployed as a
service.

A simple use case can be formulated as follows.
A user wants to find an analytics service and needs to
identify candidate services based on their descriptions and
features. A user wants to find services quickly and therefore expects
modern keyword search and taxonomy, faceted search, query
functionalities; as well as descriptions that facilitate location and
identification of relevant and appropriate analytics services, from the
registry.

The registry contains enough details to not only locate the service,
but also how to use it.

\paragraph*{Access Requirements.} 


Public Analytics Service Registry. Public analytics discovery
services are intended to allow users to find publicly hosted
services. The information provided includes the provider, [x], and
[y], and / thus reduces users' efforts in locating relevant services.

\begin{description}

\item[Levels of Assurance (LoA) in User Identity.] Most readers should
  be familiar with functionality to {\em sign in with ORCHID, or
    Facebook} or something known to the user. In general identity
  management scenarios, this provision enables what is referred to as
  {\em guest identities}, which is useful for many users who are
  interested in invoking low-level activities or less sensitive
  operations. With respect to federated service authentication and
  authorization, OIDC guest identities meet a low level of
  assurance. In contrast, users with higher LoAs are afforded
  permissions to perform to privileged activities or gain access to
  more sensitive xyz.

\item[Multi factor Authentication in User Identity.] A means for
  authenticating users via two or more types of
  authentication. An MFA instrument can elevate a user's level of
  assurance profile. RAF and IGTF are examples of such assurance
  framework standards.  OpenID Connect, SAML, and X.509 are examples
  of services that expose interfaces for multiple authentication.

\item[Private Analytics Service Registry.] Analytics Services stored
  in private registries are only available to authenticated and
  authorized / member users. Private registries allow providers to
  build virtual organizations [/ VOMS] that advertise specialized
  services to its user community. In contrast to a public analytics
  registry, access controls in private registries are more
  restricted. In addition, different group privileges may restrict the
  visible analytics service to the user (see related sections on
  user identity and levels of user privilege).

\item[Federated Analytics Service Registry.] A user wants to make
  selection decisions regarding which service to use. Analytics
  service brokers and providers therefore offer a federated analytics
  service in which multiple services from multiple providers are
  included. Rather than having to visit multiple, separate providers'
  registries, the user can visit the federated registry of the
  analytics broker to look up all potentially suitable services, via a
  single interface and browser. It may be expected that federated
  registries abstract the technical effort that casual users would
  experience during location and inspection of published analytics
  services.  Underlying analytics service registry technologies
  leverage cross-organization persistent identifiers, enhanced with
  information that the original service provider may not have
  available, and xyz. such "enrichment" may could include, for example,
  cost comparisons, or (some type of) ratings from its user community.

\item[Enhanced Analytics Service Registry.] Both public and private
  registries may need to be enhanced by providing detailed information
  so the user has a better understanding of the offering and allows
  comparison to similar artifacts maintained and published in the
  registry. Information details may include, for example, benchmark
  information, service level agreements, or cost measures such as
  carbon cost, or technical limitations such as storage access and
  availability for big data.
  
\end{description}

\paragraph*{Registry Namespace.}
To allow uniform integration of entries into a unified namespace, URLs
are used to distinguish the services. This includes two different
entities. Firstly, an entity that defines the code base of a
service. Such a code base could be for example hosted on publicly
accessible code repositories. Secondly, the namespace could include
instantiated analytics service endpoints that define a running
instance of an analytics service.

The attributes are listed in Table \ref{tab:reg}. Some attributes may
be optional and may be dependent on whether they are deployed services, or
contain a library that may be deployed.

\begin{table}[htb]
\caption{AS services Catalog and Registry attributes}\label{tab:reg}
\resizebox{1.0\columnwidth}{!}{%
\begin{tabular}{|p{3cm}|p{5cm}|p{0.25cm}|p{0.25cm}p{0.25cm}|}
\hline
&             & \rotatebox{90}{Catalog provider} & \rotatebox{90}{Service provider} & \rotatebox{90}{Library provider} \\
Name & Description & & \multicolumn{2}{l}{Register} \\
\hline
ID & 	UUID, globally unique &	\OK & \OK &	\OK \\
Name & 	Name of the service	& \OK & \OK	& \OK \\ 
Title & 	Human readable title & \OK &	\OK	& \OK \\
Public	& True if Public
(needs use case to delineate what pub private means) &  \OK &	\OK & \OK \\
Description	& Human readable short Description of the Service	& \OK & \OK & 	\OK \\
Endpoint &	The endpoint of the service	& \OK & \OK	&  \NA \\
List of Input Parameters &
	A list of parameters to the service. The parameters have each the form of name, function, type, value, and access. The type indicates the data type. The access indicates if the parameter is a data stream, database, single value/function, or event.
The function responds to a different function in case multiple are provided by the service.	& \OP & \OK	& \OK \\ 
List of Output Parameters 
  style (event, stream, data)
  value
  timestamp & 
	List of responses cast by the service. The responses have the form of function, name, type, value, access, and timestamp. The type indicates the data type. The access indicates if the parameter is a data stream, database, single value/function, or event.
The function responds to a different function in case multiple are provided by the service. &  \OP &  	\OK  & \OK \\
Version	& The version number or tag of the service	& \OK & \OK	& \OK \\
License	& The license description	& \OK & \OK	& \OK \\
Protocol & 	Example: REST	& \OK & \OK	& \OK \\
Microservice & 	True if microservices used & \OK & \OK	& \OK \\
Modified & 	Modification Timestamp	& \OK & \OK& \OK \\
Owner	& Name of the distributing entity, organization or individual. It could be a vendor.	& \OK & \OK	& \OP \\
Author &	Contact details of the people or organization responsible for the service	& \OK & \OP	& \OK \\
Tags &	Human readable common tags that are used to identify the service that are associated with the service	& \OK & \OP & \OP \\
Categories &	A category that this service belongs to (NLB, Finance, ...)	& \OK & \OP & \OP \\
Created	& date and time on which the analytics service was instantiated or created	instantiated	& \OK & \OK & \OK \\
Heartbeat &	State and timestamp of the last check when the service was active	& \NA & \OP & 	\NA \\
Documentation &	Link to a URL with a detailed description of the service
Source	Link to the source code if available	& \OK & \OP & \OP \\
Specification/Schema &	Pointer to where specification schema is located	& o & \OK &  \OK \\
AdditionalMetadata	& Pointer to where additional is located including the one here.	& \OP & \OP &	\OP \\
SLA	& Serves level agreement including cost	& \OP & \OP 	& \OP \\
CachingInterval	&If a service is accessed a lot, the caching interval can be used to put a limitation on the Response with an LRU cache	& + & \OP &	\NA \\
DataIntegration &	In case of big data the data cannot be provided as a parameter to the analysis function. Instead, we need to provide the data as endpoint. However, often tata may need to be uploaded or can be downloaded. In this case this field provides the upload and download endpoints and the protocol to access the data	& \OP & \OP &	\OP \\
Authors	& contact details of the people or organization responsible for the service (freeform string)	& \OK & \OK & \OK \\
\hline 
\multicolumn{5}{l}{\OK = required; \OP = optional, \NA = not applicable}\\
\end{tabular}
}

\end{table}

\paragraph*{Benefits of a federated analytics service registry}

A service registry can publicize and improve end-user access to data
from different sources, by overcoming some of the challenges inherent
in describing and surfacing document content and format. Publication,
and discovery of information resources are enriched with metadata
enabling the findability and reusability of a service supporting the
FAIR principle. While describing the interfaces and allowing for the
instantiation or the reuse of already instantiated services we address
the accessibility and interoperability. With respect to analytics as a
service, end users should be able to find various analytic services
and similar services without having to individually search multiple
locations or databases, each built to operate on its own, unique
storage and retrieval constructs. Through these descriptions automated
service integration can be provisioned while targeting not only the
functionality involved, but also allowing service level considerations
to be addressed. Furthermore, such analytics services could provide
significant security implications such as the protection of a database
while only exposing a subset of approved analytics functions that are
executed on the data sets. This includes partial and controlled
sharing of data mashups that can be made available to the community and
registered to make reuse easier without everyone having to replicate
the service.

%% file: section/federation.tex
\FILE{section-federation.tex}

\subsection{Service Federation}
\label{sec:federation}

This section discusses aspect of federated registries to locate and
consume analytics services with persistent identifiers across
organizations.

This is not the term is at this time in the document not properly used.

We use so far 

(1) federation of catalog and registry
(2) federation of services stored in the registry and catalog
(3) federaion of services through high level services delegatiing to other services. 


%% file: section/workflow.tex
\FILE{section-workflow.tex}

\subsubsection{Analytics Service Pipelines}

\paragraph{Motivation.}
In many cases a big data analysis is split up in multiple
subtasks. These subtasks may be reusable in other analytics
pipelines. Hence it is desirable to be able to specify and use them in
a coordinated fashion allowing reuse of the logic represented by the
analysis. Users must have a clear understanding on what the analysis
is doing and how it can be invoked and integrated.

\paragraph{Access Requirements.}
The analysis must include a clear and easy to understand specification
that encourages reuse and provides sufficient details about its
functionality, data dependency and performance. Analytics services may
have authentication, autorotation and access controls build in that
enable access buy users controlled by the service providers.

\begin{figure}[htb]
\centering\includegraphics[width=1.0\columnwidth]{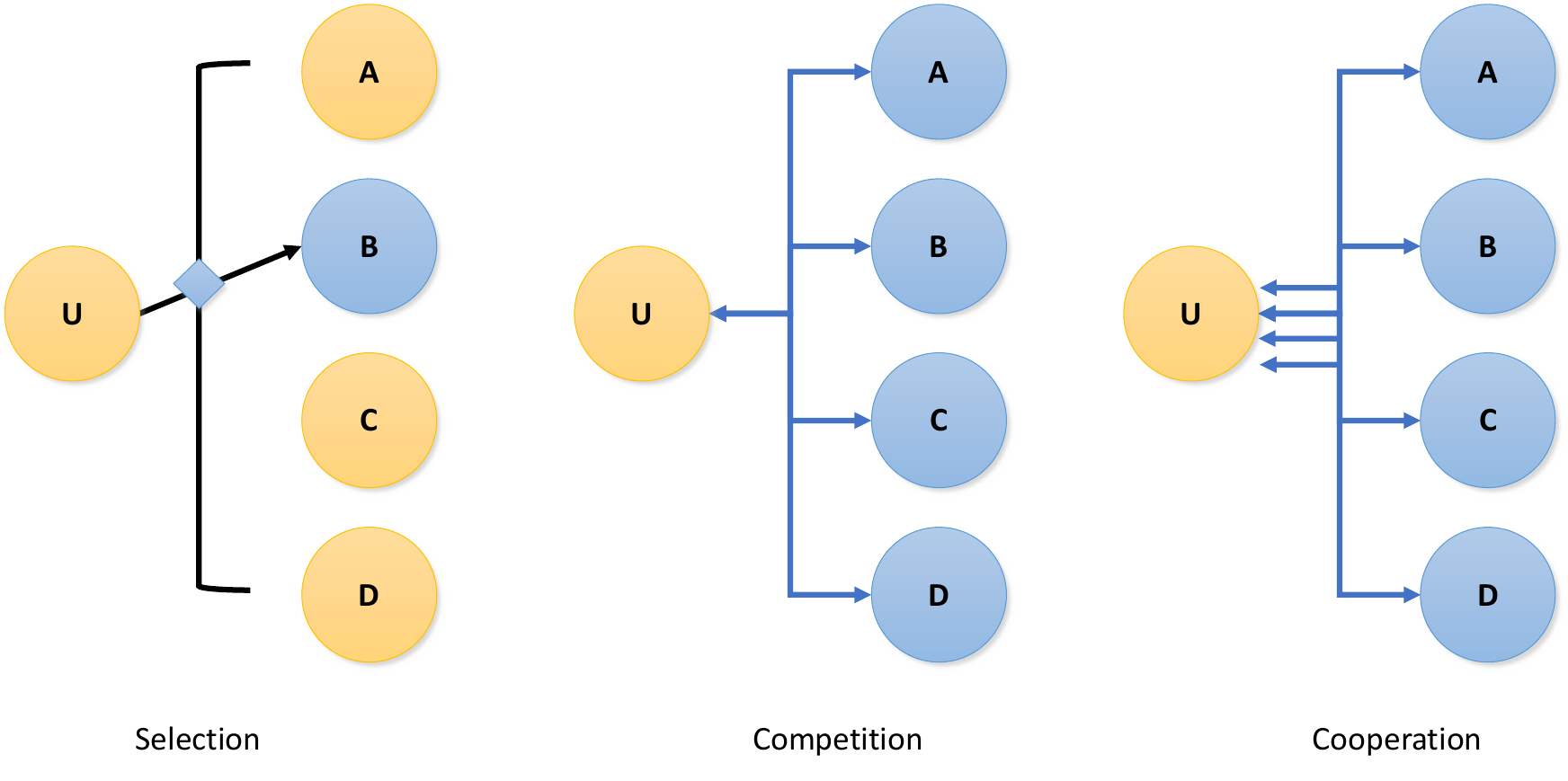}
\label{fig:hvac-2}
\caption{Hybrid service integration.}
\end{figure}

\subsubsection{Workflow Controlled Computing}

High-performance computing (HPC) is for decades a very important tool for science. Scientific tasks can leverage the processing power of a supercomputer so they can run at previously unobtainable high speeds or utilize specialized hardware for acceleration that otherwise are not available to the user. HPC can be used for analytic programs that leverage machine learning applied to large data sets to, for example, predict future values or to model current states. For such high-complexity projects, 
there are often multiple complex programs that may be running repeatedly in either competition or cooperation. This may include resources in the same or different data centers. We developed 
a hybrid multi-cloud analytics service framework that was created to manage heterogeneous and remote workflows, queues, and jobs. It can be used through a Python API, the command line, and a REST service. It is supported on multiple operating systems like macOS, Linux, and Windows 10 and 11. 
The workflow is specified via an easy-to-define YAML file.
Specifically, we have developed a library called Cloudmesh Controlled Computing (cloudmesh-cc) that adds workflow features to control the execution of jobs on remote compute resources, while at the same time leveraging capabilities provided by the local compute environments to directly interface with graphical visualizations better suited for the desktop. The goal is to provide numerous workflows that in cooperation enhances the experience of the analytics tasks. This includes a REST service and command line tools to interact with it.

\begin{figure}[htb]
\centering\includegraphics[width=0.7\columnwidth]{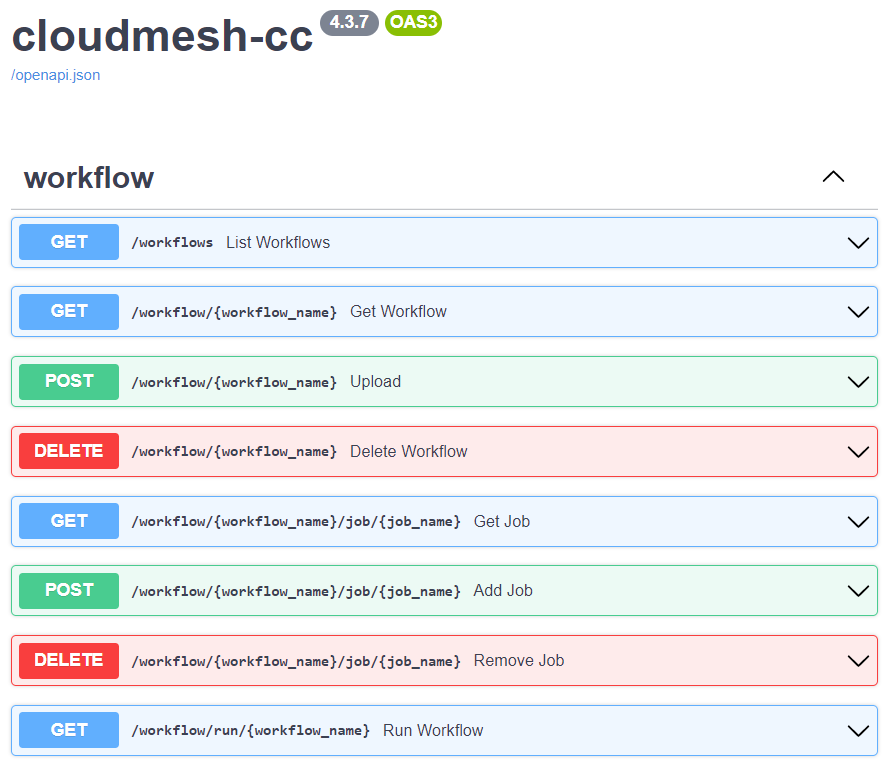}
\caption{Fast API Workflow Service.}
\label{fig:fastapi-cc-arch}
\end{figure}

\begin{figure}[htb]
    \centering
    \includegraphics[width=1.0\columnwidth]{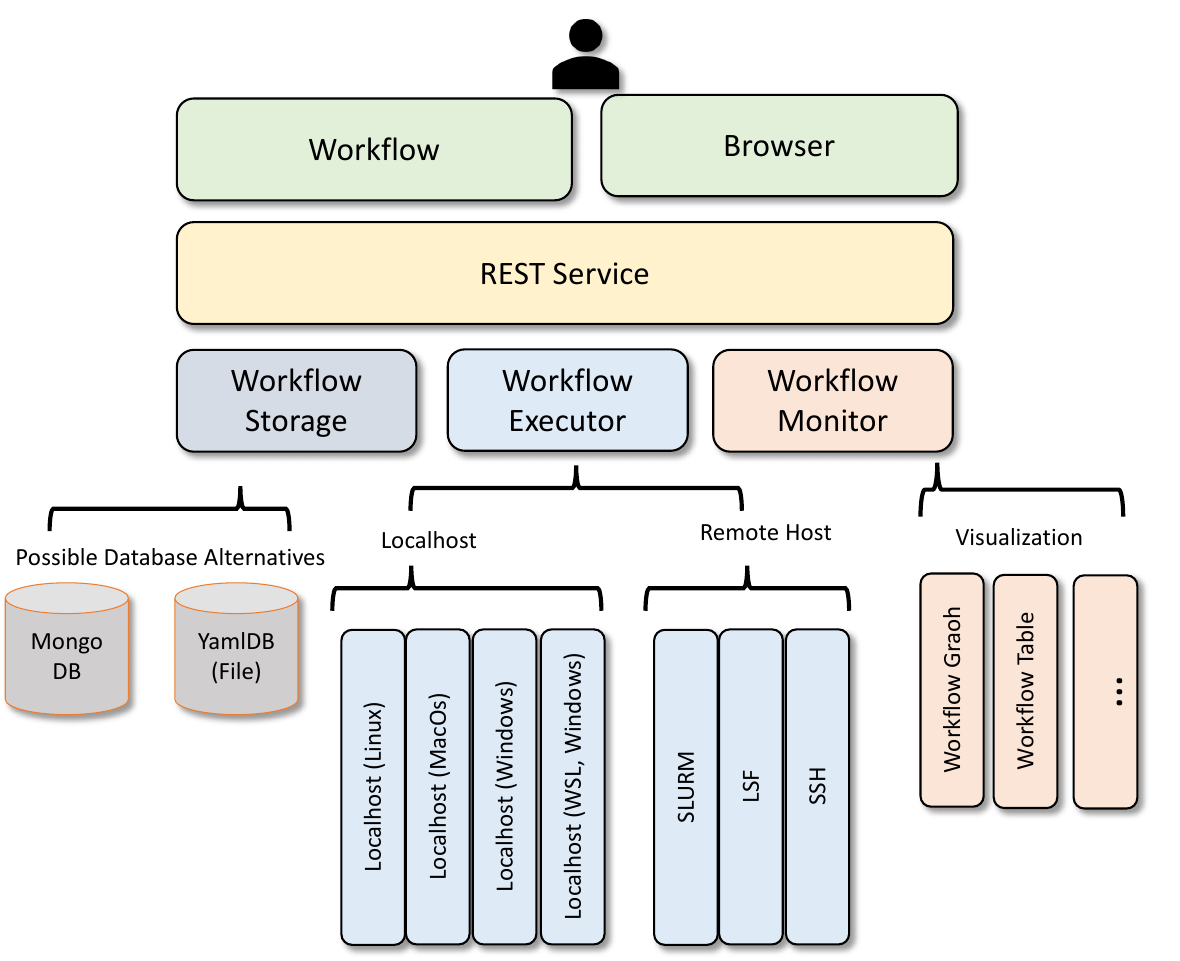}
    \caption{Architecture Workflow Service.}\label{fig:cc-2}
\end{figure}

\begin{figure}[htb]
    \centering
    \includegraphics[width=1.0\columnwidth]{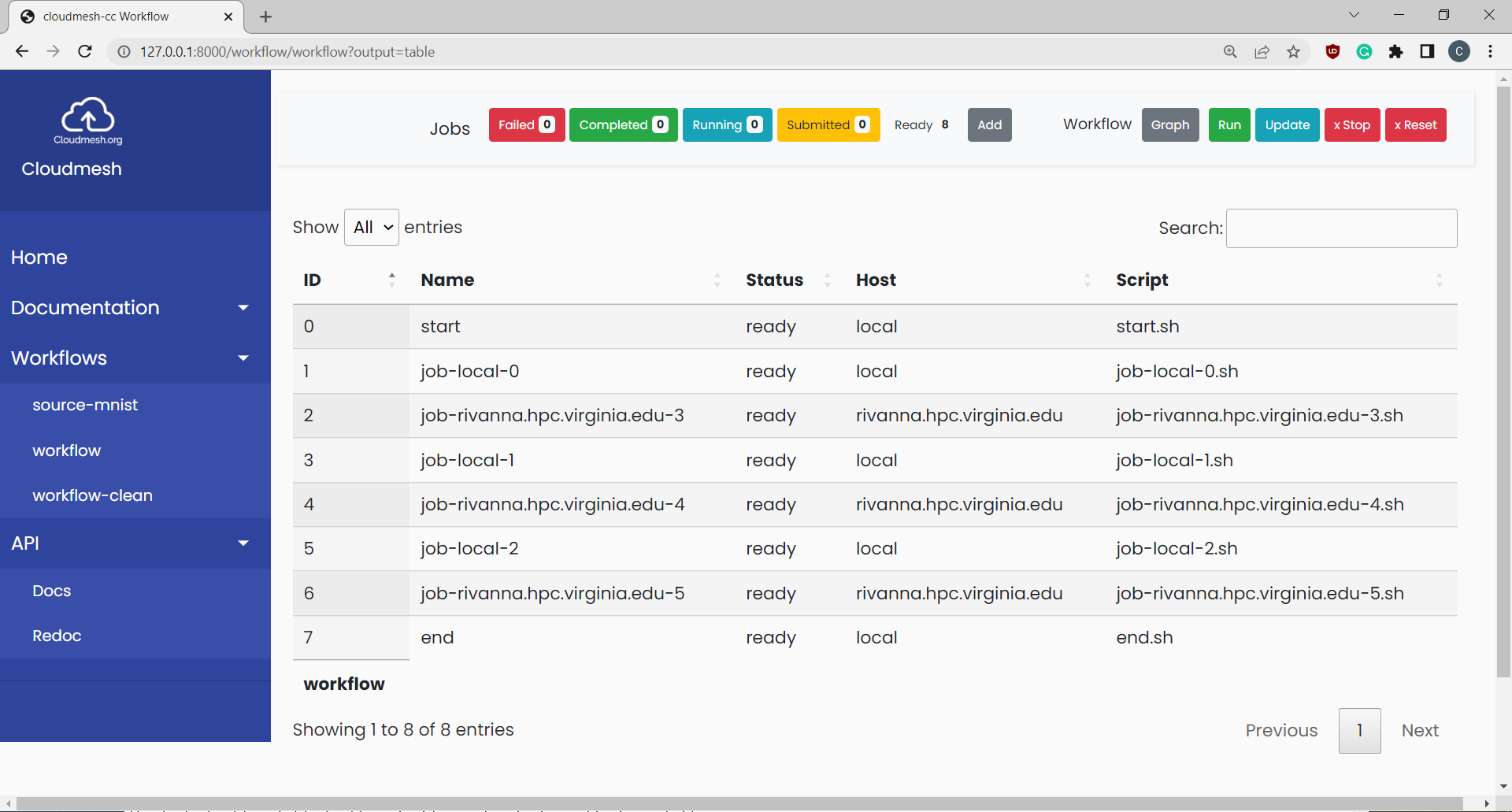}
    \caption{Workflow user interface.}\label{fig:cc-3}
\end{figure}

We have tested the framework while running various MNIST application examples, including include Multilayer Perceptron, LSTM (Long short-term memory), Auto-Encoder, Convolutional, and Recurrent Neural Networks, Distributed Training, and PyTorch training. 
A much larger application using earthquake prediction has also been used.

Figure \ref{fig:fastapi-cc-arch} shows the REST specification and Figure \ref{fig:cc-2} shows the architecture. A user interface of the running application is shown in Figure \ref{fig:cc-3}.



%% file: section/data.tex
\FILE{section-data}
\subsection{Data Management}
\label{sec:data}

As data is to be integrated and analyzed as well as produced the
framework must include a sufficiently broad ability to access data from
a wide variety of sources and utilize many different services.  This
includes files, object stores, and databases.  To address the access of
file-based data we have integrated into cloudmesh components to deal
with transfers and copies from remote resources, as well as developed
a component that deals with many file transfers through Globus. The
later overcomes restrictions due to file numbers and improves transfer
speed, and access speed.

Managing data in analytics services is an important implicit
requirement. The data has to be readily accessible and may have to be
pre-staged to the resources where the computation is performed. Also
one has to make sure that policy restrictions are appropriately dealt
with in order to perform the analytics tasks. The policy restrictions
typically include the total size of the data for a particular user or
group but also could include the number of files.

For this reason, it is advantageous to have a service that can deal with
such restrictions. Unfortunately, such services are not readily
available based on our experience with different HPC centers offering
compute resources for analytics tasks and jobs.

Available services are typically restricted to filesystems that are
accessible on the compute nodes as well as services that copy between
local computers or between compute centers. The later is frequently
covered by `rsync` over SSH or UDP, or through
Globus \cite{www-globus-transfer} as a service. However, when we tried
using Globus we found that it is not usable when millions of files are
involved, but performs well when in such cases a tar file is produced
over many files and the tar file is transfered in a single
operation. We also encountered frequent timeouts on the servers that
were involved in a server-server transfer using many file transfers.

To simplify this we developed a program
cloudmesh-globus \cite{cloudmesh-globus} that allows us to specify an
entire directory with many files that first automatically package the
directory and transfers the compressed file to the remote machine
where it then gets uncompressed and placed in the appropriate file
system. Hence such steps have not to be done by hand by the
researcher, but are done automated providing a simplified
filesystem-to-filesystem service via Globus without issues.

Other alternatives could include
cloudmesh-storage \cite{cloudmesh-storage} which include prototype
transfer services even among cloud providers such as amazon, azure,
and Google, while leveraging a compute services conducting file
copies between the involved parties.

%% file: section/package.tex
\FILE{section-package.tex}

\subsection{Package Analytic Algorithms as Service Payloads}
\label{sec:package}

It is important to package analytic algorithms with well-defined
input and output parameters as service payloads that can be reusable,
deployable, and operational across multi-cores, CPUs, and GPU
computing platforms.
In addition, it is possible to package analytics services on several 
different levels. This includes containers, binary packages, source
code, as well as the distribution of analytics functions that can be
send between services.

%% file: section/experimment.tex
\subsection{Analytics Parameter Study Experiments}

One of the important aspects of analytics as a service is the ability to integrate long-running analytics tasks either 
on the current computer, remote computers, or batch queuing systems from HPC tasks. These tasks can be 
executed directly on the host system but can simplify access for the user by placing them behind REST 
services. This is a common compute pattern as part of parameter studies that explore a variety of 
parameters producing analytics results that can then be explored either in collaboration or competition. 

For this purpose, we use two services. First, the Cloudmesh Compute Coordinator, that coordinates computational tasks onto hybrid heterogeneous resources, and second, the Cloudmesh Experiment Executor, that coordinates how to execute various parameter settings to achieve the desired results with these analytics settings.

Together these services allow the following.

\begin{enumerate}
    \item Provisioning one or multiple compute nodes on which the parameter experiment is conducted.
    \item Create hyperparameters for the analytics calculation suitable for the compute resources.
    \item Prepare the compute resources with the needed data and programs to conduct the analytics 
          functions. This can be done by copying the source to the nodes or using a GitHub repository to 
          obtain the source.
    \item Configuring the system's software to prepare for a benchmark run by installing or compiling the code 
          in a way that is best suited for the resource).
    \item Executing the analytics function and capturing the results
\end{enumerate}

This logic is captured as part of the analytics parameter experiment management and is implemented using the Cloudmesh Experiment Executor (cloudmesh-ee) utility specifically targeting execution pattern \cite{cloudmesh-ee,las-2023-escience,las-2023-mlcommons-edu-eq}.
This utility provides the ability to specify configurable parameters that perform one-to-one substitutions and a special \textit{experiments} parameter set, which creates a permutation for all parameter values as distinct experiments.

\begin{figure*}[htb]
    \centering
    \includegraphics[width=0.8\textwidth]{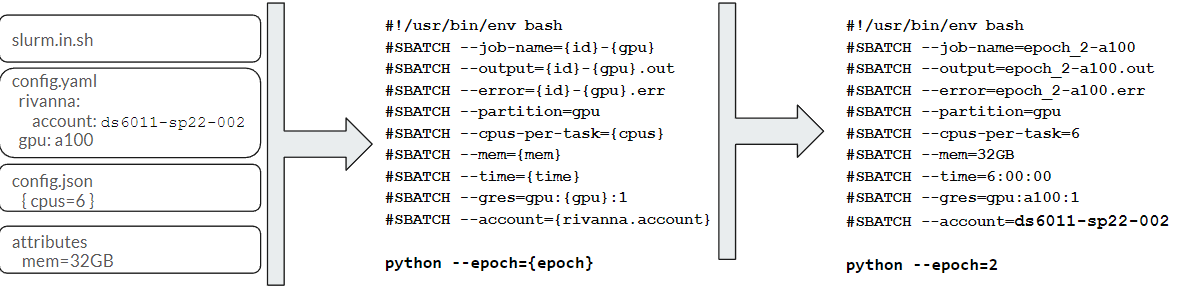}
    \caption{Compute Coordinator Generation Workflow}\label{fig:ee}

    \centering
    \includegraphics[width=0.8\textwidth]{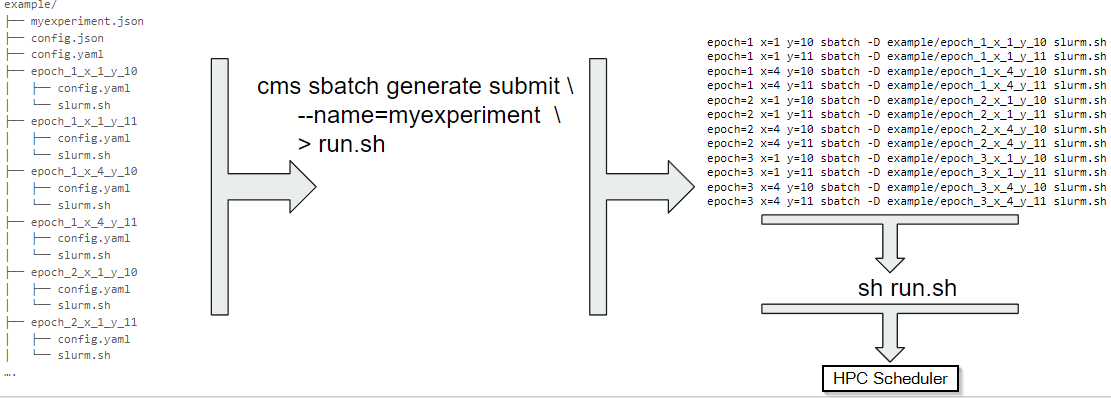}
    \caption{Compute Coordinator Submission Workflow}\label{fig:ee-submit}

    \centering
    \includegraphics[width=0.8\textwidth]{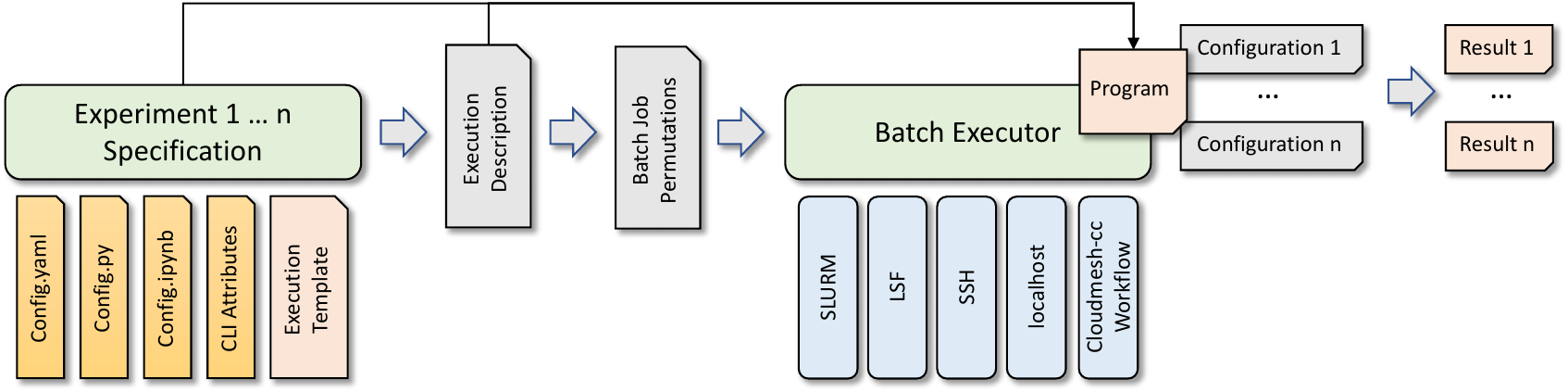}
    \caption{Compute Coordinator Sumission Workflow}\label{fig:sbatch-2}
\end{figure*}

This allows the user to provide a single script with a configuration file containing multiple hyperparameter values and have them expand into hundreds of configurations without having to prepare each configuration manually.
A typical workflow of development is illustrated in Figure \ref{fig:ee} and Figure \ref{fig:ee-submit}.

A more in-depth presentation of these services is provided in \cite{las-2023-ai-workflow,las-2023-escience}.

%% file: section/nlp.tex
\FILE{section/nlp.tex}

\subsection{NLP}

Natural Language Processing is one of the first services offered by
many cloud providers. This is motivated by analyzing large amounts of
text in volume and number and deriving automated content from
it. Popular services include keyword extraction, sentiment analysis,
auto summary, and translation.  The services are offered often by large
cloud providers such as Google, IBM Watson, and Amazon to consumers for
a fee. In addition, such tools are also offered as stand-alone
components and software packages.

As many such services are offered by the different providers, and
standalone components and software packages exist, it allows us to use
them to test the framework for implementing hybrid and multi-cloud
analytics services. We can therefore analyze each of their APIs and
compare functionality as well as the performance characteristics of
local as well as cloudservices. We also can test our design of the
cloud service catalog to identify the strategy of dynamically
locating similar services and integrating them into a service offering.

For our work, we have restricted our analysis to two cloud services
from Amazon and Google, while the integration of a third from Amazon
is under development. Furthermore, we have only considered the
translation service as it provides an easy abstraction of a service
that translates a text from a source language to a target language:

\smallskip
\begin{Verbatim}[fontsize=\small]
def translate(text, source_langauge, target_langauage, ...):
    ...
    return translation
\end{Verbatim}
\smallskip

Each of the services is implemented with a different API. We contrast
the API in Figure ... showcasing the difference in invoking a
translation service as well as showcasing the result of the JSON
response of such a service.

If the interface is on purpose defined differently a switch will cost
extra work and may therefore not be in the interest of the users.  It
is obvious that users can benefit from a uniform implementation of
this API in order to easily switch from one provider to the other.
Naturally, the cloud providers typically do not have any interest in
providing such a uniform API as it may entice the customers to switch
service providers.

Hence a multicloud CLI implementation may look as follows, where the
provider flag is used to distinguish the different cloud providers
offering the translation service. Naturally, we could also utilize a
local translation programs such as offered by industry and easily make
this example is a hybrid service that also integrates with a local
implementation.

\begin{Verbatim}[fontsize=\small]
cms nlp translate --provider=google --from=en --to=de hello world
cms nlp translate --provider=aws --from=en --to=de hello world
cms nlp translate --provider=local --from=en --to=de hello world
\end{Verbatim}

As each of the services returns natively a different output, it is
beneficial to unify the output and create a mapping from the
originating service to the output. An example of such a uniform output is given next.

\begin{Verbatim}[fontsize=\small]
{'date': '05/02/2022 14:45:45',
 'input': 'hello world',
 'input_language': 'en',
 'output': 'Hallo Welt',
 'output_language': 'de',
 'provider': 'aws',
 'time': 0.2641}
\end{Verbatim}

Other parameters such as service region can easily be integrated in
this example. Furthermore, it is obvious that the commandline
application underlying API can be used in a REST service
implementation and can be generalized into different REST service
frameworks. For our implementation, we have used FastAPI and used the
closest regional service center to our location.

\begin{figure}[htb]

\centering
\includegraphics[width=1.0\columnwidth]{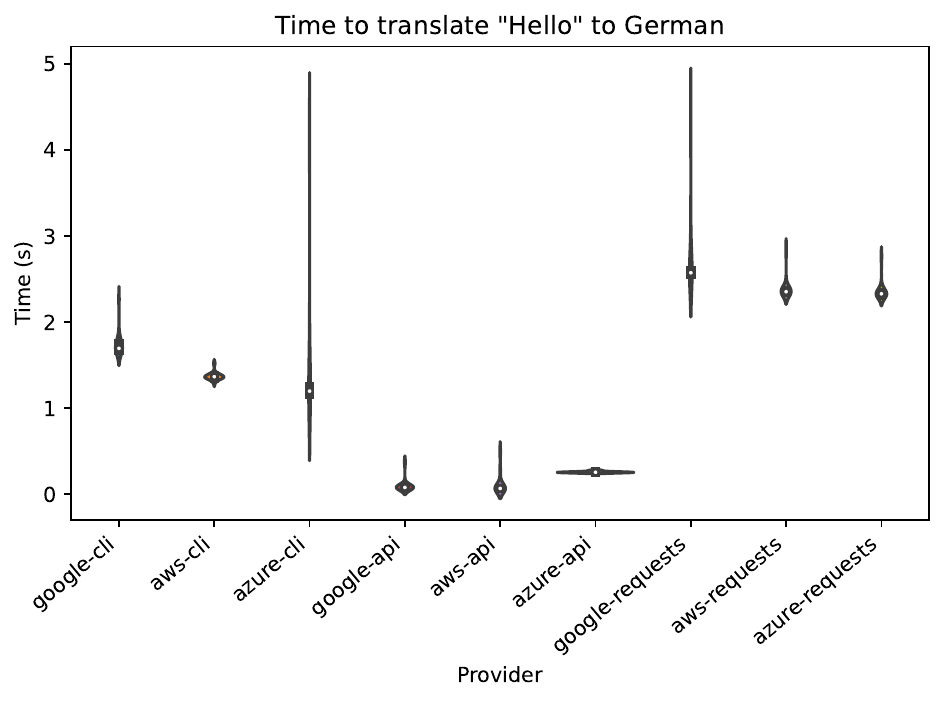}

\caption{Natural Language processing performance for a single
         word translation.}
\label{fig:nlp-performance}

\end{figure}

The result of the translation that simply translates a text from
German ``Hallo Welt'' to English ``Hello World'' is showcased for 100
invocations in Figure~\ref{fig:nlp-performance}.

Such a performance analysis could be performed based on customer needs
and could indicate factors for preferential service choices. This may
include besides time other factors such as cost and quality, or even
service level reliability at different times.

For us, we found that for the short query, we used the service
offered by AWS to translate the text was on average 
faster when executed from Bloomington, IN to the closest service
centers for the provider.

\begin{table*}[htb]

\caption{Comparision of the time it takes on average to translate the word hello from English to German}

\begin{tabular}{lrrrrrrrrr}
 & google-cli & aws-cli & azure-cli & google-api & aws-api & azure-api & google-requests & aws-requests & azure-requests \\
 \hline
count & 20.000000 & 20.000000 & 20.000000 & 20.000000 & 20.000000 & 20.000000 & 20.000000 & 20.000000 & 20.000000 \\
mean & 1.737850 & 1.369900 & 1.365600 & 0.094800 & 0.099300 & 0.257150 & 2.681500 & 2.381500 & 2.352550 \\
std & 0.134381 & 0.043240 & 0.664569 & 0.066135 & 0.104365 & 0.009292 & 0.429593 & 0.111200 & 0.100339 \\
min & 1.643000 & 1.297000 & 1.120000 & 0.069000 & 0.064000 & 0.245000 & 2.532000 & 2.330000 & 2.299000 \\
25\% & 1.675250 & 1.346750 & 1.162750 & 0.075000 & 0.066750 & 0.251000 & 2.556750 & 2.349250 & 2.316500 \\
50\% & 1.695000 & 1.366500 & 1.198500 & 0.079000 & 0.067500 & 0.255000 & 2.574000 & 2.354000 & 2.329500 \\
75\% & 1.749000 & 1.378750 & 1.246750 & 0.085750 & 0.069250 & 0.259500 & 2.592750 & 2.367750 & 2.341500 \\
max & 2.266000 & 1.520000 & 4.170000 & 0.374000 & 0.497000 & 0.280000 & 4.481000 & 2.848000 & 2.766000 \\
\hline
\end{tabular}
\end{table*}

%% file: section/eq-result.tex
\FILE{section/eq-result.tex}

\subsection{Earthquake Result}

\begin{figure*}[p]
     \centering
     \begin{subfigure}[b]{0.49\textwidth}
        \centering\includegraphics[width=1.0\linewidth]{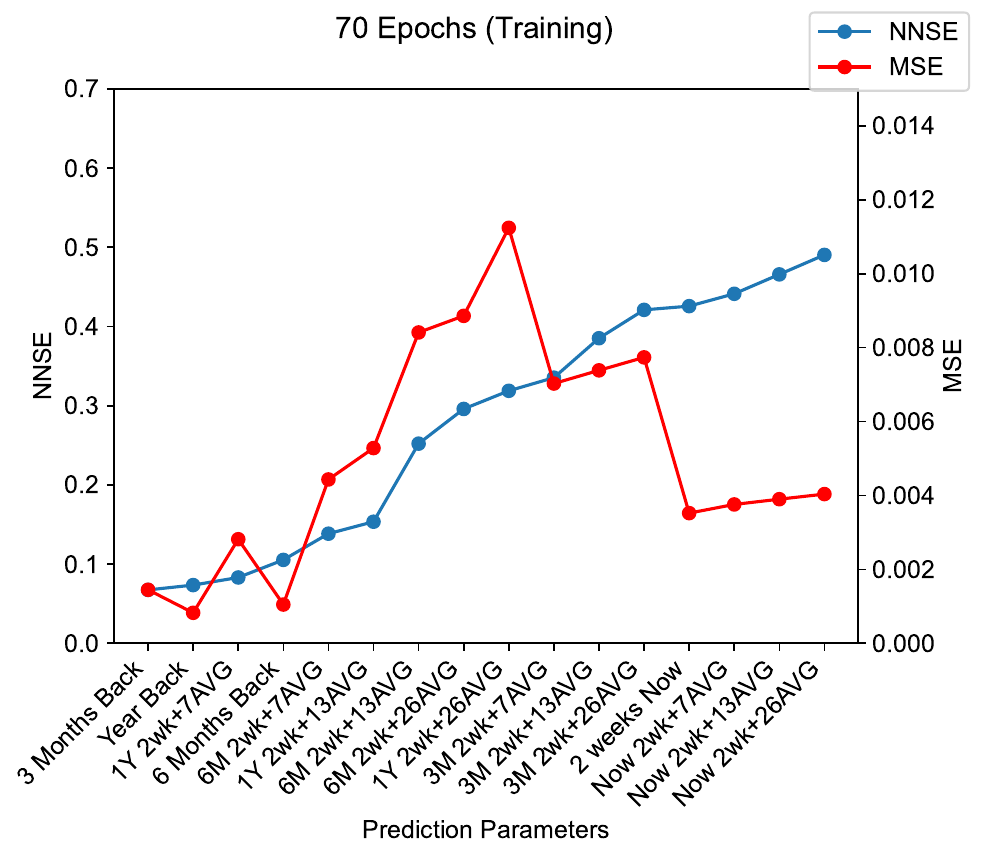}
        \caption{MSE and NNSE - 70 epochs training}
        \label{fig:eq1}
     \end{subfigure}
     \hfill
     \begin{subfigure}[b]{0.49\textwidth}
        \centering\includegraphics[width=1.0\linewidth]{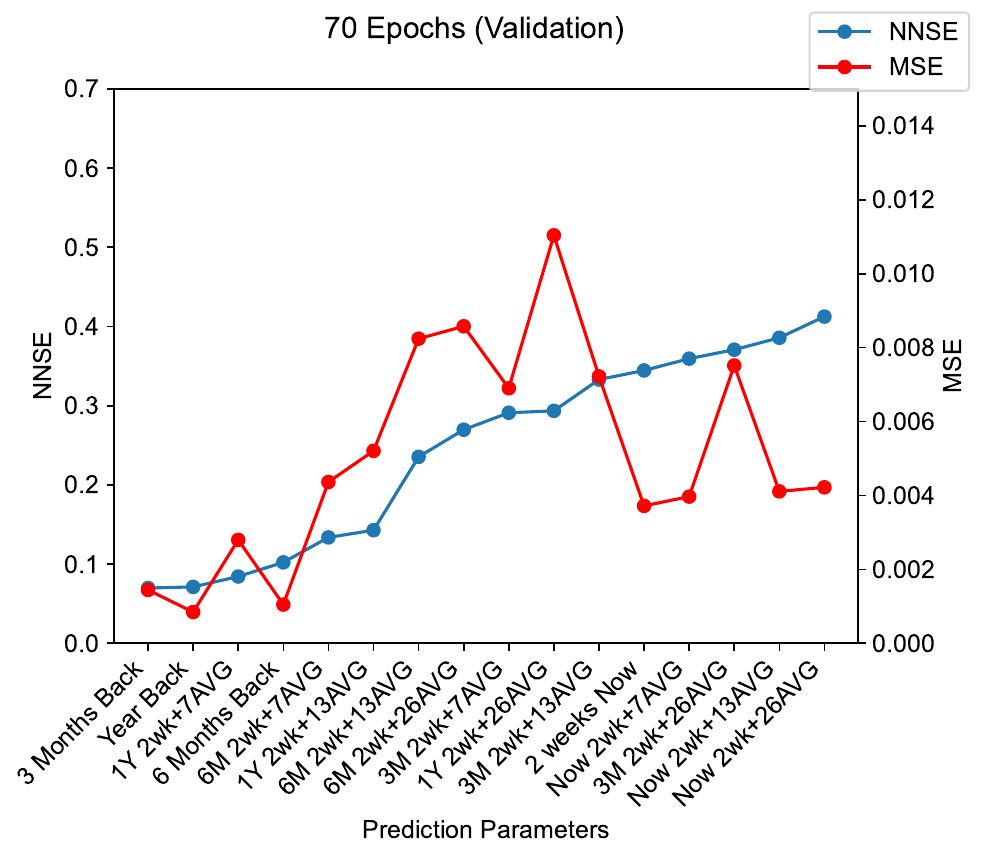}
        \caption{MSE and NNSE - 70 epochs validation}
        \label{fig:tbd4}
     \end{subfigure}
        \caption{NNSE and MSE values for 70 epochs 70 (training and validation).}
        \label{fig:eq2}
\end{figure*}

While using this framework we have conducted a parameter study for earthquake forecasting. Figure \ref{fig:eq1} and \ref{fig:eq2} show the results of this application. A more in depth paper about this application has been published in \cite{las-2023-mlcommons-edu-eq}, \cite{las-2023-escience} and \cite{las-2023-ai-workflow}.

%% file: section/conclusion.tex
\section{Conclusion}

In this paper, we have outlined our thoughts on building an analytics services framework. We have integrated aspects to make these services hybrid services across various compute providers including HPC, clouds, and microservices.

In addition to this exploratory thought, a component-based prototype implementation leveraging cloudmesh has been used to verify the feasibility of the approach. It has then been tested theoretically on some applications, while it also has been practically explored in the area of natural language processing and earthquake forecasting.

Furthermore, the framework has been used intensely as part of the MLCommons Science Working group where it was used for additional application.

%% file: NIST-acknowledgement.tex
\section*{Acknowledgement}

Work was in part funded by NIST 60NANB21D151T  the NSF 1829704 and 2200409.  The work was also funded by the Department of Energy under the grant Award No. DE-SC0023452. The work was conducted at the Biocomplexity Institute and Initiative at the University of Virginia.

This manuscript has been co-authored by UT-Battelle, LLC under Contract No. DE-AC05-00OR22725 with the U.S. Department of Energy. The publisher, by accepting the article for publication, acknowledges that the U.S. Government retains a non-exclusive, paid up, irrevocable, world-wide license to publish or reproduce the published form of the manuscript, or allow others to do so, for U.S. Government purposes. The DOE will provide public access to these results in accordance with the DOE Public Access Plan (http://energy.gov/downloads/doe-public-access-plan).